\documentclass[journal]{IEEEtran}

\IEEEoverridecommandlockouts
\usepackage{cite}
\usepackage{amsmath,amssymb,amsfonts}
\usepackage{algorithmic}
\usepackage{graphicx}
\usepackage{textcomp}
\usepackage{xcolor}
\usepackage{mathtools}
\usepackage{ntheorem}
\usepackage{epstopdf}
\usepackage{caption}
\usepackage{setspace}
\usepackage{subcaption}
\usepackage{dsfont}
\usepackage{cleveref}
\usepackage{mathrsfs}
\newcommand*{\rom}[1]{\expandafter\@slowromancap\romannumeral #1@}
\usepackage{romannum}

\makeatletter
\renewcommand*\env@matrix[1][*\c@MaxMatrixCols c]{%
\hskip -\arraycolsep
  \let\@ifnextchar\new@ifnextchar
  \array{#1}}
\makeatother

\newcommand{\Real}{{\mathds R}}

\newcommand{\Nat}{{\mathds N}}

{}
\newtheorem{corollary}{Corollary}{}
\newtheorem{proposition}{Proposition}{}
{}
\newtheorem{theorem}{Theorem}{}
\newtheorem{remark}{Remark}{}
\newtheorem{lemma}{Lemma}{}
{}

\begin{document}
%
\title{\LARGE \bf
Privacy-Preserving Anomaly Detection in Stochastic Dynamical Systems: Synthesis of Optimal Gaussian Mechanisms
}
%
%

\author{Haleh Hayati, Nathan van de Wouw, Carlos Murguia%
\thanks{Haleh Hayati, Nathan van de Wouw, and Carlos Murguia are with the Department of Mechanical Engineering, Dynamics and Control Group, Eindhoven University of Technology, The Netherlands. Emails: \& h.hayati@tue.nl, \& n.v.d.wouw@tue.nl, \& c.g.murguia@tue.nl.}
}

\markboth{Journal of \LaTeX\ Class Files,~Vol.~14, No.~8, August~2015}%
{Shell \MakeLowercase{\textit{et al.}}: Bare Demo of IEEEtran.cls for IEEE Journals}
\maketitle


\begin{abstract}
We present a framework for designing distorting mechanisms that allow remotely operating anomaly detectors while preserving privacy. We consider the problem setting in which a remote station seeks to identify anomalies using system input-output signals transmitted over communication networks. However, disclosing true data of the system operation is not desired as it can be used to infer private information -- modeled here as a system private output. To prevent accurate estimation of private outputs by adversaries, we pass original signals through distorting (privacy-preserving) mechanisms and send the distorted data to the remote station (which inevitably leads to degraded monitoring performance). We formulate the design of these mechanisms as a privacy-utility trade-off problem. 
We cast the synthesis of dependent Gaussian mechanisms as the solution of a convex program where we seek to maximize privacy quantified using information-theoretic metrics (mutual information and differential entropy) over a finite window of realizations while guaranteeing a bound on monitoring performance degradation.
\end{abstract}
\IEEEpeerreviewmaketitle
\section{Introduction}
In today's hyperconnected world, the rapid advancement of science and technology has resulted in massive user data being collected and processed by numerous companies over public networks. Companies use this data to provide personalized services and targeted advertising. However, these technologies have come at the cost of a significant loss of privacy. Adversaries with sufficient resources can infer sensitive (private) information about users/systems from public data available on the internet and/or unsecured networks/servers. This is why researchers from different fields 
have been drawn to privacy and security of Cyber-Physical Systems (CPSs), \cite{Farokhi1}-\nocite{Farokhi2}\nocite{Jerome1}\nocite{ferrari2021safety}\cite{Carlos_Iman1}.

In most engineering applications, information about the state of systems is obtained through sensor measurements and sent to a remote station through communication networks for signal processing and decision-making. One of these applications is remote anomaly detection, where a remote station seeks to identify anomalies based on system data transmitted over communication networks.  However, if the communication network is public/unsecured and/or the remote station is not trustworthy, adversaries might access and estimate systems' private data. To avoid this, we propose a distorting mechanism to randomize sensor and input data before transmission using additive \emph{dependent} Gaussian random vectors designed to hide the private part of the state. The aim of this work is to devise synthesis tools to design these mechanisms such that private information is protected and, at the same time, the implementation of remote anomaly detectors is still possible using distorted data. Therefore, the synthesis tool must consider the trade-off between privacy and monitoring performance. We remark that we protect against both eavesdropping adversaries and the (potentially untrustworthy) remote station. We aim to develop a privacy scheme that considers no trusted party and seeks to destroy information in the direction of private outputs as much as possible while keeping monitoring performance degradation bounded.

In this manuscript, we follow an information-theoretic approach to privacy. As \emph{privacy metric}, we propose a combination of \emph{mutual information} and \emph{entropy} \cite{Cover} between disclosed and private data. To characterize anomaly detection performance, we consider the false alarm rate of standard chi-squared detectors as \emph{performance metric}. Hence, the synthesis algorithm seeks to optimize the privacy metric by selecting the stochastic properties of the injected dependent noise (their joint covariance matrices over a finite window of realizations) while not degrading the monitoring false alarm rate by more than a desired amount. We prove that this problem can be posed as a constrained convex optimization problem -- log-determinant cost with linear matrix inequality constraints.

Using additive random noise is common practice to enforce privacy of sensitive data. Differential Privacy (DP) is a popular method in the context of privacy of dynamical systems \cite{Jerome1,koufogiannis2017differential}. However, in most cases, privacy is defined at every time-step, i.e., they do not consider information disclosed over a history of observations. While this gives some privacy guarantees, it is well known that for dynamical systems, the more data is collected, the more accurate the input/state estimation can be. For instance, in the deterministic case, if the system is observable and of dimension $n$, after $n$ time-steps, the state can be completely reconstructed without error. Therefore, if the privacy scheme is to avoid input/state estimation, giving guarantees at every time-step is certainly not sufficient. Moreover, injecting i.i.d. noise at every time-step \cite{ferrari2021differentially,yazdani2022differentially} is also conservative as there is an obvious correlation among measurements and inputs in dynamical systems.

There are existing techniques to address privacy in dynamical systems from an information-theoretic perspective \cite{Topcu,murguia2021privacy,Farokhi2}. In these methods, privacy is characterized using information-theoretic metrics -- e.g., mutual information, entropy, and Fisher information. However, independently of the metric being used, if the data to be kept private follows continuous probability distributions, the problem of finding the optimal additive noise to maximize privacy is difficult to solve \cite{Farokhi1}. This issue has been addressed by assuming the data to be kept private is deterministic \cite{Farokhi1}. But, in a CPSs context, the inherent system dynamics and unavoidable system and sensor noise lead to stochastic non-stationary data, and thus, existing tools do not fit.
In \cite{hayati2021finite,hayati2022gaussian}, we have started exploring these ideas assuming Gaussianity in system disturbances and without considering any particular application at the remote station -- we mainly focused on formulating the privacy-utility problem over a finite horizon and characterized utility degradation by the expected squared distortion between distorted and original data. Motivated by these results, in this manuscript, we propose a general class of Gaussian mechanisms and seek the optimal parametrization of the additive dependent noise distribution (its joint covariance matrix over a window of realizations) given the system dynamics and the anomaly detector at the remote station.

\indent To the best of the authors' knowledge, there is only one related result in the literature. For remote anomaly detection in stochastic dynamical systems, the authors in \cite{ferrari2021differentially} consider differential privacy and provide tools to select the variance of i.i.d. additive noise so that disclosed data guarantees a certain level of differential privacy. Their results consider privacy of input signals (so no private state-dependent outputs), and at every time-step, i.e., they do not consider information disclosed over a history of observations. Although these results provide a DP privacy guarantee on input data, the use of i.i.d. noise at every time-step leads, in general, to conservative results as there is an obvious correlation between measurements and inputs in dynamical systems that must be accounted for to maximize privacy. Moreover, results in \cite{ferrari2021differentially} do not provide a mathematical characterization of privacy-detection tradeoffs nor provide synthesis methods to design optimal distorting mechanisms (in the sense of optimizing these tradeoffs).
Taking into account these points, our contributions are:
\begin{itemize}
\item The characterization of proper privacy and detection metrics tailored to the specific application domain (private remote anomaly detection in stochastic dynamical systems) in terms of system matrices. This allows quantifying (enforcing) given (desired) trade-off between privacy and anomaly detection performance.
\item A synthesis framework (in terms of a series of convex programs) to design additive-dependent Gaussian mechanisms over finite horizons that minimize information leakage while guaranteeing a desired anomaly detection performance. To accomplish this, we have reformulated the original design optimization problem (which is non-convex in cost and constraints) as a log-determinant cost with linear matrix inequality constraints for the lifted system dynamics.
\item We prove that our framework can handle prior distributions in the system dynamics that are not necessarily Gaussian, as long as they are log-concave \cite{prekopa1980logarithmic}. The latter further generalizes the class of stochastic dynamical systems that our framework can handle. Due to space limitations, the latter is included in an extended pre-print of this paper \cite{hayati2022privacy}.
\end{itemize}  

\subsection{Notation}
The symbol $\Real$ stands for the real numbers. The symbol $\Nat$ stands for the set of natural numbers. The Euclidian norm in $\Real^n$ is denoted by $||X||$, $||X||^2=X^{\top}X$, where $^{\top}$ denotes transposition. The $n \times n$ identity matrix is denoted by $I_n$ or simply $I$ if $n$ is clear from the context. Similarly, $n \times m$ matrices composed of only ones and only zeros are denoted by $\mathbf{1}_{n \times m}$ and $\mathbf{0}_{n \times m}$, respectively, or simply $\mathbf{1}$ and $\mathbf{0}$ when their dimensions are clear. The notation $\operatorname{col}\left(x_1, \ldots, x_n\right)$ stands for the column vector composed of the elements $x_1, \ldots, x_n$. This notation will also be used in case the components $x_i$ are vectors. For positive definite (semidefinite) matrices, we use the notation $P>0$ ($P \geq 0$). 
For any two matrices $A$ and $B$, the notation $A \otimes B$ (the Kronecker product) stands for the matrix composed of submatrices $A_{ij}B$, where $A_{ij}$, $i,j=1,\ldots,n$, stands for the $ij-$th entry of the $n \times n$ matrix $A$. 
The notation $X \sim \mathcal{N}[\mu,\Sigma^X]$ means $X \in \Real^{n}$ is a normally distributed random vector with mean $E[X] = \mu \in \Real^{n}$ and covariance matrix $E[(X-\mu)(X-\mu)^T] = \Sigma^X \in \Real^{n \times n}$, where $E[a]$ denotes the expected value of the random vector $a$. The notation $Y \sim \Gamma(K, \theta)$ means $Y$ follows a gamma distribution with shape parameter $K$ and scale parameter $\theta$. Finite sequences of vectors are written as $X^K := (X(1)^{\top},\ldots,X(K)^{\top})^{\top} \in \Real^{Kn}$, $X(i) \in \Real^{n}$, $i \in \{1,\ldots,K\}$, and $n,K \in \Nat$. To avoid confusion, we denote powers of matrices as $(A)^{K} = A \cdots A$ ($K$ times) for $K > 0$, $(A)^{0} = I$, and $(A)^{K} = \mathbf{0}$ for $K < 0$. The operators $\log(\cdot)$, $\det(\cdot)$, and $\text{tr}(\cdot)$ are logarithm, determinant, and trace.
\section{System Description and Anomaly Detection}
\subsection{System Description}
We consider discrete-time stochastic systems of the form:
\begin{eqnarray}\label{eq1}
\left\{ \begin{aligned}
x_{k+1} &= Ax_k + Bu_k + t_k + G \delta_k,\\
y_k &= Cx_k + w_k + H \delta_k,\\
s_k &= D x_k,
\end{aligned} \right.
\end{eqnarray}
with time-index $k \in \Nat$, state $x_k \in {\mathbb{R}^{{n_x}}}$, measurable output $y_k \in {\mathbb{R}^{{n_y}}}$, \emph{known} input $u_k \in {\mathbb{R}^{{n_u}}}$, private performance output $s_k \in {\mathbb{R}^{{n_s}}}$, and matrices $(A,B,C,D,G,H)$ of appropriate dimensions. Vector $\delta_k \in {\mathbb{R}^{{n_\delta}}}$ models changes in the system dynamics due to faults or anomalies (thus, in the absence of anomalies $\delta_k=\mathbf{0}$).
Matrix $D$ is full row rank, and the pair $(A,C)$ is detectable. The state disturbance $t$ and the output disturbance $w$ are multivariate i.i.d. Gaussian processes with zero mean and positive definite covariance matrices ${\Sigma ^t}$ and ${\Sigma ^w}$, respectively. The initial state $x_1$ is assumed to be a Gaussian random vector with $E[x_1]=\mu^x_1 \in \mathbb{R}^{n_x}$ and covariance matrix $\Sigma^x_1 \in \mathbb{R}^{n_x \times n_x}$, $\Sigma^x_1 > 0$. Disturbances $t_k$ and $w_k$ and the initial condition $x_1$ are mutually independent. We assume that matrices (vectors) $(A,B,C,D,\Sigma^x_1,\mu^x_1,\Sigma^t,\Sigma^w)$ and input $u_k$ are known for all $k$. 
\subsection{Anomaly Detection}
The aim of the anomaly detection algorithm is to identify anomalies (i.e., when $\delta_k \neq \mathbf{0}$ in \eqref{eq1}) using system input-output signals. Here, we consider standard Kalman filter-based chi-squared change detection procedures \cite{murguia2019model}. The main idea is to use an estimator to anticipate the system evolution in the absence of anomalies. This prediction is subsequently compared with actual measurements from sensors. If the difference between what is measured and the estimation (commonly referred to as residuals) is larger than expected, there might be an anomaly in the system. We use steady-state Kalman filters as the state estimator.
\subsubsection{Steady-state Kalman Filter (anomaly-free case)}
Consider the following one step-ahead Kalman filter \cite{Astrom} for \eqref{eq1}:
\begin{equation}\label{eq5b}
\hat{x}_{k+1} = A \hat{x}_k + Bu_k + L (y_k - C\hat{x}_k),
\end{equation}
with estimated state $\hat{x}_k \in {\mathbb{R}^{{n_x}}}$, $\hat{x}_1 = E[x_1]$, and output injection gain matrix $L \in {\mathbb{R}^{{n_x} \times {n_y}}}$. Define the estimation error $e_k := x_k - \hat{x}_k$. The optimal filter gain $L$ minimizing the trace of the asymptotic estimation error covariance matrix $P := \lim_{k \rightarrow \infty} E[e_k e_k^\mathrm{T}]$ in the absence of anomalies is given by
\begin{equation}
    L:=\left(A P C^\top\right)\left(\Sigma^w + C P C^\top\right)^{-1},
\end{equation}
where $P$ is the solution of the Riccati equation:
\begin{equation}
    A P A^\top-P+ \Sigma^t=A P C^\top\left(\Sigma^w +C P C^\top\right)^{-1} C P A^\top.\label{solveP}
\end{equation}
Equation \eqref{solveP} always has a unique solution $P$ under the assumption of detectability of the pair $(A,C)$ \cite{Astrom}.

\begin{figure}[!t]
  \includegraphics[width=3.2in]{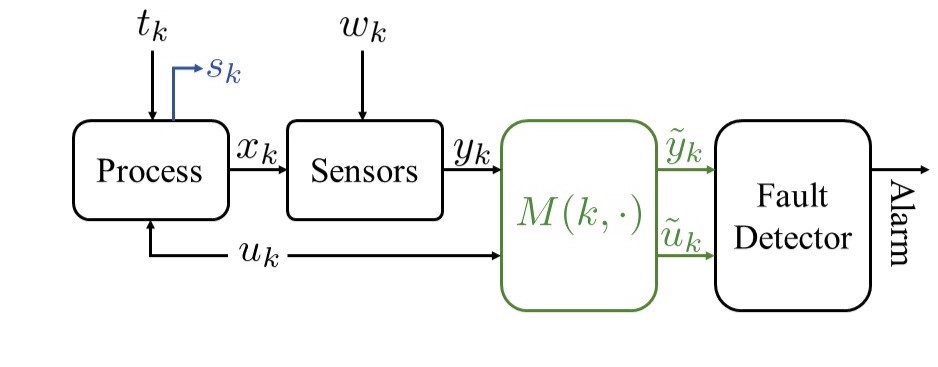}
  \caption{System configuration.}
    \label{fig1}
\end{figure}
\subsubsection{Residuals and Hypothesis Testing}
Consider the process dynamics \eqref{eq1} and the steady-state Kalman filter \eqref{eq5b}, and define the residual sequence $r_k := y_k - C\hat{x}_k$, which can be shown to evolve according to the difference equation:
\begin{align}\label{eq7b}
\left\{
\begin{aligned}
    e_{k+1} &= (A-L C) e_k + t_k - L w_k + (G-LH) \delta_k, \\
    r_k &= C e_k + w_k + H \delta_k.
\end{aligned}
\right.
\end{align}
If there are no anomalies ($\delta_k=\mathbf{0}$), the mean and the covariance matrix of the residual are as follows:
\begin{align}
\left\{
\begin{aligned}
    &E[r_{k}] = C E[e_k] + E[w_k]= \mathbf{0}_{n_y \times 1}, \\[2mm]
    &E[r_{k} r^{\top} _{k}]=CPC^\top +\Sigma^w =: \Sigma \in {\mathbb{R}^{{n_y} \times {n_y}}}.
\end{aligned}\label{eq9}
\right.
\end{align}
For this residual, we identify two hypotheses to be tested: $\mathcal{H}_0$ the normal mode (no anomalies) and $\mathcal{H}_1$ the faulty mode (with anomalies). Under the normal mode, $\mathcal{H}_0$, the statistics of the residual is given by \eqref{eq9}.
However, in the presence of anomalies ($\mathcal{H}_1$), we expect that the statistics of the residual are different from those in the normal mode, i.e.,
\begin{equation}\label{eq10}
\mathcal{H}_1:\left\{
\begin{aligned}
    &E[r_{k}] \ne  \mathbf{0}_{n_y \times 1}, \text{ and/or}\\[2mm]
    &E[r_{k} r^{\top} _{k}] \ne \Sigma.
\end{aligned}
\right.
\end{equation}

There exist many well-known hypothesis testing techniques that could be used to examine the residual and subsequently detect anomalies/faults. 
In this manuscript, we consider a particular case of CST, namely the so-called chi-squared change detection procedure \cite{murguia2016cusum,IET_CARLOS_JUSTIN}.
\subsubsection{Distance Measure and Chi-Squared Procedure}
The input to the chi-squared procedure is a distance measure $z_k \in {\mathbb{R}}$, i.e., a measure of how deviated the estimator is from the sensor measurements. We use the quadratic distance measure
\begin{equation} \label{distance}
    z_k = r^\top _k \Sigma^{-1} r_k,
\end{equation}
where $r_k$ and $\Sigma$ are the residual sequence and its covariance matrix introduced in \eqref{eq9}. If there are no anomalies ($\delta_k=\mathbf{0}$), i.e. $E[r_{k}] = \mathbf{0}$ and $E[r_{k} r^{\top} _{k}]= \Sigma$, it follows that
\begin{equation}\label{eq11}
\left\{
\begin{array}{ll}
    E[z_{k}] = \text{tr}[\Sigma ^{-1} \Sigma]+ E[r_k]^\top \Sigma ^{-1} E[r_k] = n_y,\\[2mm]
    \begin{aligned}
            \text{var}[z_{k}]&=  2\text{tr}[\Sigma ^{-1} \Sigma \Sigma ^{-1} \Sigma]+ 4E[r_k]^\top \Sigma ^{-1} \Sigma \Sigma ^{-1} E[r_k] \\[1mm]
    &= 2n_y,
        \end{aligned}
\end{array}
\right.
\end{equation}
see, e.g. \cite{ross2014introduction} for details. Because $r_k \sim  \mathcal{N}(\mathbf{0},\Sigma)$, $z_k$ follows a chi-squared distribution with $n_y$ degrees of freedom. The chi-squared procedure is characterized by $z_k$ and its cumulative distribution as follows:\\[1mm]
\noindent\makebox[\linewidth]{\rule{\linewidth}{0.8pt}}
\textbf{Chi-Squared procedure:}\\ 
If $z_{k}=r_{k}^{\top} \Sigma^{-1} r_{k}>\alpha$, $\tilde{k}=k$.\\
\textbf{Design parameter:} threshold $\alpha \in \mathbb{R}_{+}$. \\
\textbf{Output:} alarm time(s) $\tilde{k}$.\\
\noindent\makebox[\linewidth]{\rule{\linewidth}{0.8pt}}\\
The idea is that if $z_k$ exceeds the threshold $\alpha$, alarms are triggered, and the time instant $k$ is identified as an alarm time (denoted by $\tilde{k}$). The parameter $\alpha$ is selected to satisfy a desired false alarm rate $\mathcal{A}^*$. Assume that there are no anomalies and consider the chi-squared procedure with threshold $\alpha \in \mathbb{R^+}$, $r_k \sim  \mathcal{N}(\mathbf{0},\Sigma)$. Let $\alpha=\alpha^{*}:=2 {P}^{-1}\left(\frac{n_y}{2}, 1-\mathcal{A}^{*}\right)$, where ${P}^{-1}(\cdot,\cdot)$ denotes the inverse regularised lower incomplete gamma function, then the false alarm rate $\mathcal{A}$, induced by the threshold $\alpha$, satisfies $\mathcal{A} = \mathcal{A}^*$ (see \cite{IET_CARLOS_JUSTIN} for proof and details).
\subsection{Privacy Mechanism and Adversarial Capabilities}
We aim to prevent adversaries from accurately estimating the private output $s_k$ (see \eqref{eq1}). To this end, we randomize measurements $y_k$ and input signals $u_k$ before transmission and send the corrupted data to the remote station instead. The idea is to randomize $y_k$ and $u_k$ through a stochastic mapping $M(k,y_k,u_k)$ of the form:
\begin{equation}\label{eq2}
M(k,y_k,u_k) := \left\{
\begin{array}{ll}
  \tilde{y}_k = y_k + v_k,\\[2mm]
  \tilde{u}_k = u_k + j_k,
\end{array}
\right.
\end{equation}
for some Gaussian processes, $v_k \sim \mathcal{N}[\mathbf{\mathbf{0}},\Sigma^v_k]$ and $j_k \sim \mathcal{N}[\mathbf{\mathbf{0}},\Sigma^j_k]$ with zero mean and time-varying covariance matrices $\Sigma^v_k$ and $\Sigma^j_k$ to be designed for each time instant. We consider $v_k$ and $j_k$ as dependent Gaussion processes, implying that privacy noises sequences, ${v^K} = \operatorname{col}(v_1,\ldots,v_K)$ and ${j^K} = \operatorname{col}(j_1,\ldots,j_K)$ for $\mathcal{K}:=\{1,\ldots,K\}$ are dependent sequences. The randomized vectors $\tilde{y}_k$ and $\tilde{u}_k$ are disclosed to the remote station to be used by the remote anomaly detector to detect faults/attacks and trigger alarms, see Figure \ref{fig1}. 
Since at the remote station, only the received distorted signals $(\tilde{y}_k,\tilde{u}_k)$ can be used, the dynamics of the system used at the remote station in the presence of anomalies and privacy distortions is as follows:
\begin{eqnarray}\label{eq4}
\left\{ \begin{aligned}
x_{k+1} &= Ax_k + B\tilde{u}_k +  t_k+G\delta_k \\&= Ax_k + Bu_k + B j_k +  t_k+G\delta_k,\\
\tilde{y}_k &= y_k + v_k = Cx_k + w_k+H\delta_k+ v_k.
\end{aligned} \right.
\end{eqnarray}
That is, the original plant driven by $\tilde{u}_k$ and producing $\tilde{y}_k$.\\
We consider worst-case adversaries that eavesdrop data at the communication network and/or the remote station. They do not only have access to all distorted sensor measurements $\tilde{y}_k$ and distorted input signals $\tilde{u}_k$, but also have prior knowledge of the dynamics and the stochastic properties of the system, i.e., matrices $(A,B,C,D,\Sigma^x_1,\mu^x_1,\Sigma^t,\Sigma^w)$ are known by the adversary. Moreover, the adversary also knows the means and covariance matrices of $\tilde{y}_k$, $(\mu^{\tilde{y}}_k,\Sigma^{\tilde{y}}_k)$, and $\tilde{u}_k$, $(\mu^{\tilde{u}}_k,\Sigma^{\tilde{u}}_k)$, as these can be estimated from the disclosed data $(\tilde{y}_k,\tilde{u}_k)$. We assume that the adversary uses a linear MMSE estimator (see \cite{huemer2017component}) to reconstruct the private output $s_k$, which, for jointly Gaussian vectors, produces the best estimation performance among all unbiased estimators \cite{huemer2017component}. In practice, actual adversaries would typically not have all the capabilities that we assume here. However, if we maximize privacy under such worst-case adversaries, we ensure that adversaries with less capabilities perform even worse (or equal at most).
\subsection{Performance Degradation Due to Privacy Mechanisms}
Note that whenever the privacy mechanism in \eqref{eq2} is in place, the anomaly detector at the remote station uses the distorted disclosed data $(\tilde{y}_k,\tilde{u}_k)$ instead of $(y_k,u_k)$ to drive its Kalman filter and subsequently detect anomalies. That is, the remote filter takes the following form:
\begin{equation}\label{eq5}
\hat{x}_{k+1} = A \hat{x}_k + B\tilde{u}_k + L (\tilde{y}_k - C\hat{x}_k).
\end{equation}
Because we assume that the gain matrix $L$ has been designed a priori without considering the privacy-preserving mechanism (so considering the stochastic properties of $(y_k,u_k)$ and not $(\tilde{y}_k,\tilde{u}_k)$), the use of distorting mechanisms in \eqref{eq2} will inevitably lead to anomaly detection performance degradation.

Consider the system \eqref{eq4} and the distorted Kalman filter \eqref{eq5}, and define the distorted residual sequence $\tilde{r}_k := \tilde{y}_k - C\hat{x}_k$, which can be shown to evolve according to the difference equation:
\begin{equation}\label{eq7}
\left\{
\begin{array}{ll}
\begin{aligned}
    \tilde{e}_{k+1}&= (A-L C) \tilde{e}_k + t_k - B j_k - L v_k - L w_k \\
    &+ (G-LH)\delta_k,\end{aligned}
 \\[2mm]
    \tilde{r}_k= C \tilde{e}_k + w_k + v_k +H \delta_k.
\end{array}
\right.
\end{equation}
Hence, if we apply the privacy filter \eqref{eq2}, the statistics of the distorted residual $\tilde{r}_k$ in \eqref{eq7} in the anomaly-free case ($\delta_k=\mathbf{0}$) are:
\begin{equation}\label{eq12}
\left\{
\begin{array}{ll}
    \begin{aligned}
    &E[\tilde{r}_{k}] = CE[\tilde{e}_k]+E[w_k]+E[v_k]=\mathbf{0}_{n_y \times 1}, \\
    &E[\tilde{r}_{k} \tilde{r}^{\top} _{k}] = C E[\tilde{e}_{k} \tilde{e}^{\top} _{k}] C^\top+ \Sigma^w  + \Sigma^v_k\\
    &\quad= C(P + B\Sigma^j_k B^\top + L\Sigma^v_k L^\top )C^\top + \Sigma^w  + \Sigma^v_k\\
    &\quad=\Sigma + \Sigma^v_k + C L\Sigma^v_k L^\top C^\top + CB\Sigma^j_k B^\top C^\top=: \tilde{\Sigma}_k.
    \end{aligned}
\end{array}
\right.
\end{equation}
Then, by applying the privacy filter to the measurement and input, the residual becomes a 
multivariate Gaussian process with covariance matrix $\tilde{\Sigma}_k$ as in \eqref{eq12}. It follows that the distance measure used for anomaly detection takes the form $\tilde{z}_{k} = \tilde{r}^{\top}_{k} \Sigma^{-1} \tilde{r}_{k}$ in the distorted case. Note that because we assume the detector is designed without considering the privacy mechanism, the distorted distance measure $\tilde{z}_{k}$ is still normalized by the undistorted residual covariance metric $\Sigma$ introduced in \eqref{eq9}. This results in $\tilde{z}_{k}$ not following a chi-squared distribution anymore. Therefore, the detection threshold $\alpha$ (which was computed for the undistorted measure $z_k$) will always lead to performance degradation when applied to the distorted $\tilde{z}_{k}$ for detection.
\section{Privacy-utility trade-off Formulated as an Optimization Problem}
As discussed in previous sections, adding the privacy mechanism changes the residual properties, see \eqref{eq12}, which results in increasing the false alarm rate. The privacy noise in \eqref{eq2} acts as an additional source of uncertainty from the detection point of view, which may reduce the performance of the anomaly detection scheme. We aim to formulate an optimization problem to characterize a desired trade-off between privacy and the performance of the anomaly detection algorithm. For a given time horizon $\mathcal{K}:=\{1,\ldots,K\}$, $K\in \mathbb{N}$, the aim of the proposed privacy scheme is to make the inference of the sequence of private output vectors, ${s^K} =\operatorname{col}(s_1,\ldots,s_K)$, from the distorted disclosed sequences, ${{\tilde{y}}^K} = \operatorname{col}({\tilde{y}}_1,\ldots,{\tilde{y}}_K)$ and \linebreak ${{\tilde{u}}^{K}} = \operatorname{col}({\tilde{u}}_1,\ldots,{\tilde{u}}_K)$, by designing privacy noises sequences, ${v^K} = \operatorname{col}(v_1,\ldots,v_K)$ and ${j^K} = \operatorname{col}(j_1,\ldots,j_K)$, as difficult as possible without degrading the anomaly detection performance excessively.\\
As privacy metric, an intuitive candidate to use is the mutual information between private and disclosed data, i.e., $I[s^K;{\tilde{y}}^K,{\tilde{u}}^{K}]$. However, because the input sequence $u^{K}$ is deterministic and $v^K$ and $j^K$ are independent, it is easy to verify that $I[s^K;\tilde{y}^K,\tilde{u}^{K}] = I[s^K;\tilde{y}^K]$ (see\cite{Cover} for proof and details). That is, in the proposed setting, the randomized input data $\tilde{u}^{K}$ does not affect $I[s^K;\tilde{y}^K,\tilde{u}^{K}]$ at all. To overcome this, we use instead differential entropy \cite{Cover}. Differential entropy is a common metric in information-theoretic privacy approaches \cite{sankar2013utility,molloy2021smoothing}. It quantifies the average uncertainty in a random vector, and increasing the entropy of a random variable can make its estimation harder. Therefore, we add to $I[s^K;\tilde{y}^K]$ the negative differential entropy of the difference of $u^K$ and $\tilde{u}^K$, $-h[\tilde{u}^{K}-u^{K}] = -h[j^{K}]$, to capture the uncertainty between original and disclosed input data. That is, we propose $I[s^K;\tilde{y}^K] - h[j^{K}]$ as privacy metric to be minimized.\\
To capture performance degradation in the anomaly detection algorithm, we consider the change in the original false alarm rate $\mathcal{A}^*$ induced by the privacy mechanism. Let $\alpha$ denote the threshold designed in the undistorted case to ensure a false alarm rate of $\mathcal{A}^*$ for the chi-squared detector. Then, we seek privacy mechanisms that do not change the false alarm rate by more than an allowed $\epsilon$ when the original $\alpha$ and the distorted residuals $\tilde{r}_k$ are used to run the detector, i.e., we seek to satisfy the following probabilistic constraint:
\begin{eqnarray}
\operatorname{Pr}[ \tilde{z}_k  > \alpha]< \mathcal{A}^* + \epsilon,
\end{eqnarray}
which is equivalent to
\begin{eqnarray} \label{constraint_a}
\operatorname{Pr}[\tilde{z}_k  \le \alpha] > 1 - \mathcal{A}^*- \epsilon,
\end{eqnarray}
where $\epsilon$ is the false alarm distortion level. Using the definition of Cumulative Distribution Function (CDF), we can write \eqref{constraint_a} as
\begin{equation}
 F_{\tilde{z}_k}(\alpha) > 1 - \mathcal{A}^* - \epsilon, \label{constraint}
\end{equation}
where $F_{\tilde{z}_k}(\alpha)$ is the CDF of $\tilde{z}_k$, i.e., $F_{\tilde{z}_k}(\alpha)=\operatorname{Pr}[\tilde{z}_k \le \alpha]$.  Therefore, the trade-off optimization problem for maximizing privacy while considering the above constraint on the false alarm rate can be written as follows:\\[2mm]
\textbf{Problem 1} Given the system dynamics \eqref{eq1}, $k \in \mathcal{K}$ with $\mathcal{K}=\{1,\ldots,K\}$, $K\in \mathbb{N}$, desired false alarm rate $\mathcal{A}^*$, maximum distortion level ${\epsilon} \in {\mathbb{R}^+}$, the chi-squared procedure threshold $\alpha$, the distorting mechanism \eqref{eq2}, and the remote Kalman filter \eqref{eq5}, 
find the sequence of covariance matrices ${\Sigma^v_k}$ and ${\Sigma^j_k}$ in \eqref{eq2} as the solution to the following optimization problem:
\begin{equation}
\left\{\begin{aligned}
	&\min_{{\Sigma^v_k},{\Sigma^j_k}, k \in \mathcal{K}}\ I[{s^K};{\tilde{y}^K}] - h[{j}^{K}],\\[1mm]
    &\hspace{8mm}\text{s.t. }
     F_{\tilde{z}_k}(\alpha)> 1 - \mathcal{A}^*   - \epsilon, \\[1mm]
    &\hspace{8mm}\text{and } (v^K,j^K,y^K) \text{ mutually independent}.
\end{aligned}\right. \label{optimizationproblem}
\end{equation}
\section{Solution to Problem 1}
To solve the optimization problem \eqref{optimizationproblem}, we first need to write the cost function and constraints in terms of our design variables $\Sigma^v_k$ and $\Sigma^j_k$, $k \in \mathcal{K}$.
\subsection{Cost Function: Formulation and Convexity}
We first formulate the cost function, $I[{s^K};{\tilde{y}^K}] - h[{j}^{K}]$, in terms of the design variables $\Sigma^v_k$ and $\Sigma^j_k$. The differential entropy $h[{j}^{K}]$ is fully characterized by the covariance matrix of ${j}^{K}$, $\Sigma^j_K := E[j^K {j^K}^\top]$. Then, $h[{j}^{K}]$ is given by \cite{Cover}
\begin{equation}\label{hUprim}
h[{j}^{K}] = \frac{1}{2}\log \det \left( \Sigma^j_K \right) + \frac{Kn_u}{2} + \frac{Kn_u}{2}\log(2\pi).
\end{equation}
The mutual information $I\left[ {{s^K};{\tilde{y}^K}} \right]$ can be written in terms of differential entropies as $I[{s^K};{\tilde{y}^K}] = h[s^K]+ h[\tilde{y}^K]  - h[s^K,\tilde{y}^K]$, see \cite{Cover}. Moreover, these entropies are fully characterized by the covariance matrices of the corresponding random vectors \cite{Cover}. So, to write $I\left[ {{s^K};{\tilde{y}^K}} \right]$ in terms of $\Sigma^v_k$, we need to write the covariance matrices of $s^K$ and $\tilde{y}^K$, and their joint covariance in terms of $\Sigma^v_k$. By lifting the system dynamics \eqref{eq1} over $\mathcal{K}$, the stacked vector $((\tilde{y}^K)^\top,(s^K)^\top)^\top \in \mathbb{R}^{K(n_y + n_s)}$ can be written as
\begin{align}\label{stackedXY}
&\begin{bmatrix} \tilde{y}^K \\ s^K \end{bmatrix} = \begin{bmatrix} \tilde{C}_K  \\ \tilde{D}_K \end{bmatrix}F_K x_1 + \begin{bmatrix} \tilde{C}_K  \\ \tilde{D}_K \end{bmatrix} J_K t^{K-1}\\
&\hspace{15mm} + \begin{bmatrix}  \tilde{C}_K  \\ \tilde{D}_K \end{bmatrix} N_K u^{K-1} + \begin{bmatrix} I \\ \mathbf{0} \end{bmatrix}(w^{K}+v^{K}), \notag
\end{align}
with sequence of inputs ${u^{K-1}} = (u_1^\top,\ldots,u_{K-1}^\top)^\top$, stacked matrices $N_K := J_K(I_{K-1} \otimes B)$, $\tilde{C}_K := I_K \otimes C$, $\tilde{D}_K := I_K \otimes D$, and
\begin{equation}\label{stackedZ}
\left\{ \begin{aligned}
  F_K &:= \begin{bmatrix} I & A^\top & \hdots & (A^\top)^{K-1}  \end{bmatrix}^\top,\\
  J_K &:= \begin{bmatrix} \mathbf{0} & \mathbf{0} & \mathbf{0} & \cdots & \mathbf{0} \\ I & \mathbf{0} & \mathbf{0} & \cdots & \mathbf{0} \\ A & I & \mathbf{0} & \cdots & \mathbf{0} \\ \vdots & \vdots & \vdots & \ddots & \vdots \\[1mm] (A)^{K-2} & (A)^{K-3} & (A)^{K-4} & \cdots & I  \end{bmatrix}.
\end{aligned} \right.
\end{equation}
Let ${{\Sigma }^v_K} \in \mathbb{R}^{K n_y \times K n_y}$ and ${{\Sigma }^j_K} \in \mathbb{R}^{K n_u \times K n_u}$ denote the non-diagonal covariance matrix of the stacked additive dependent vectors $v^K$ and $j^K$. In the following lemma, we give a closed-form expression of the joint density of $(\tilde{y}^K,s^K)$.
\begin{lemma}\label{stackedDist}
\[ \begin{psmallmatrix} \tilde{y}^K \\ s^K  \end{psmallmatrix} \sim \mathcal{N}\left[\mu^{\tilde{y},s}_K,\Sigma^{\tilde{y},s}_K\right],\] with mean $\mu^{\tilde{y},s}_K \in \mathbb{R}^{K(n_s + n_y)}$ and covariance matrix $\Sigma^{\tilde{y},s}_K \in \mathbb{R}^{K(n_s + n_y) \times K(n_s + n_y)}$, $\Sigma^{\tilde{y},s}_K>0$\emph{:}
\begin{eqnarray}\label{muZS}
\mu^{\tilde{y},s}_K &:= \begin{bsmallmatrix}  \tilde{C}_K  \\ \tilde{D}_K \end{bsmallmatrix}F_K \mu^x_1 + \begin{bsmallmatrix}  \tilde{C}_K  \\ \tilde{D}_K \end{bsmallmatrix} N_K u^{K-1},
\end{eqnarray}
\begin{equation}\label{stackeconmatrix3}
\begingroup
\renewcommand*{\arraycolsep}{2pt}
\begin{aligned}
\Sigma^{\tilde{y},s}_K &:= \begin{bsmallmatrix} I \\ \mathbf{0} \end{bsmallmatrix}(I_K \otimes \Sigma^w) \begin{bsmallmatrix} I\\ \mathbf{0} \end{bsmallmatrix}^\top + \begin{bsmallmatrix} I \\ \mathbf{0} \end{bsmallmatrix}{{\Sigma }^v_K}\begin{bsmallmatrix} I \\ \mathbf{0} \end{bsmallmatrix}^\top\\[1mm]
&+ \begin{bsmallmatrix} \tilde{C}_K  \\ \tilde{D}_K \end{bsmallmatrix} F_K \Sigma^x_1 F_K^\top \begin{bsmallmatrix}  \tilde{C}_K  \\ \tilde{D}_K \end{bsmallmatrix}^\top\\[1mm]
&+ \begin{bsmallmatrix} \tilde{C}_K  \\ \tilde{D}_K \end{bsmallmatrix}J_K (I_{K-1} \otimes \Sigma^t)J_K^\top \begin{bsmallmatrix} \tilde{C}_K  \\ \tilde{D}_K \end{bsmallmatrix}^\top.
\end{aligned}
\endgroup
\end{equation}
\end{lemma}
\emph{\textbf{Proof}}: To simplify notation, we introduce the stacked vector ${\Theta ^K} := {( {{{( {{\tilde{y}^K}})}^\top},{{( {{s^K}})}^\top}} )^\top}$. By assumption, the initial condition $x_1$, and the processes, $t_k$, $w_k$, and $v_k$, $k \in \mathbb{N}$, are mutually independent, and $x_1 \sim \mathcal{N} [ {\mu _1^x,\Sigma _1^x} ]$, $t_k \sim \mathcal{N} [ {\mathbf{0},\Sigma^t} ]$, $w_k \sim \mathcal{N} [{\mathbf{0},\Sigma^w} ]$, $v_k \sim \mathcal{N} [ {\mathbf{0},\Sigma^v} ]$, for some positive definite covariance matrices $\Sigma _1^x$, $\Sigma^t$, $\Sigma^w$, and $\Sigma^v$. Then, see \cite{Ross} for details, we have ${L_1}x\left( 1 \right) \sim \mathcal{N} [ {{L_1}\mu _1^x,{L_1}\Sigma _1^xL_1^\top} ]$, ${L_2}t^{K-1} \sim \mathcal{N} [\mathbf{0}, {{L_2}({I_{K - 1}} \otimes {\Sigma ^t}){L_2}^T} ]$, ${L_3}w^K \sim \mathcal{N} [\mathbf{0}, {{L_3}({I_K} \otimes {\Sigma ^w}){L_3}^\top} ]$, ${L_4}v^K \sim \mathcal{N} [\mathbf{0}, {{L_4}{{\Sigma }^v_K}{L_4}^\top} ]$, for any deterministic matrices $L_j$, $j = 1, 2, 3, 4$, of appropriate dimensions. It follows that the stacked vector ${\Theta ^K}$ given in \eqref{stackedXY} is the sum of a deterministic vector, ${( {\tilde{C}_K^\top,\tilde{D}_K^T} )^\top}{N_K}{u^{K - 1}}$, and four independent normally distributed vectors. Therefore, ${\Theta ^K}$ follows a multivariate normal distribution with $E[ {{\Theta ^K}} ] = \mu^{\tilde{y},s}_K$ as in \eqref{muZS}.
By inspection, using the expression of ${\Theta ^K}$ in \eqref{stackedXY}, mutual independence among $x_1$, $t_k$, $w_k$, and $v_k$, $k \in N$, and the definition of $\Sigma _1^x$, $\Sigma _1^x = E[ {\left( {x_1 - \mu _1^x} \right){{\left( {x_1 - \mu _1^x} \right)}^\top}} ]$, it can be verified that the covariance matrix of ${\Theta ^K}$, $E[ {( {{\Theta ^K} - E( {{\Theta ^K}} )} ){{( {{\Theta ^K} - E( {{\Theta ^K}} )} )}^\top}} ]$, is given by $\Sigma^{\tilde{y},s}_K$ in \eqref{stackeconmatrix3}. It remains to prove that the distribution of ${\Theta ^K}$ is not degenerate, that is, $\Sigma^{\tilde{y},s}_K > 0$. Note that $\Sigma^{\tilde{y},s}_K$ in \eqref{stackeconmatrix3} can be written as
\begin{align}
\Sigma _K^{\tilde{y},s} &= \left[ {\begin{array}{*{20}{c}}
\Sigma _K^{\tilde{y}} & { \tilde{C}_K Q  \tilde{D}_K^\top}\\
{ \tilde{D}_K Q  \tilde{C}_K^\top }&{ \tilde{D}_K Q  \tilde{D}_K^\top}
\end{array}} \right],\label{CovZS}\\[2mm]
\Sigma _K^{\tilde{y}} &= { \tilde{C}_K Q \tilde{C}_K^\top +  (I_K \otimes \Sigma^w) + {{\Sigma }^v_K}},
\end{align}
with $Q = {F_K}\Sigma _1^xF_K^\top + {J_K}( {{I_{K - 1}} \otimes {\Sigma ^t}} )J_K^\top$. 
Necessary and sufficient conditions for $\Sigma _K^{\tilde{y},s}>0$ are its right-lower block is positive definite (${\tilde{D}_K Q  \tilde{D}_K^\top}>0$) and that the Schur complement of block ${\tilde{D}_K Q  \tilde{D}_K^\top}$ of $\Sigma _K^{\tilde{y},s}$, denoted as $\Sigma _K^{\tilde{y},s} / ({\tilde{D}_K Q  \tilde{D}_K^\top})$, is positive definite (\cite{zhang2006schur}, Theorem 1.12).
The right-lower block is positive definite if $\tilde{D}_K$ is full row rank and $Q$ is positive definite. Because $D$ is full row rank by assumption, matrix $\tilde{D}_K = \left( {{I_K} \otimes D} \right)$ is also full row rank (\cite{Horn2}, Theorem 4.2.15).
Note that $Q$ can be factored as follows
\begin{eqnarray}\label{Q}
Q = \left[ {\begin{array}{*{20}{c}}
{{F_K}}&{{J_K}}
\end{array}} \right]\underbrace {\left[ {\begin{array}{*{20}{c}}
{\Sigma _1^x}&\mathbf{0}\\
\mathbf{0}&{{I_{K - 1}} \otimes {\Sigma ^t}}
\end{array}} \right]}_{Q'}\left[ {\begin{array}{*{20}{c}}
{{F_K}}\\
{{J_K}}
\end{array}} \right].
\end{eqnarray}
That is, $Q$ is a linear transformation of the block diagonal matrix ${Q'}$ above. By inspection, it can be verified that matrix $P = [F_K \hspace{1mm} J_K]$, see \eqref{stackedZ}, is lower triangular with identity matrices on the diagonal; thus, $P$ is invertible. It follows that $Q = PQ'P^\top$ is a congruence transformation of ${Q'}$ \cite{boyd1994linear}. Hence, $Q$ is positive definite if and only if the block diagonal matrices of ${Q'}$ are positive definite \cite{boyd1994linear}. Matrices ${\Sigma _1^x}$ and ${\Sigma^t}$ are positive definite by assumption (which implies ${{I_{K - 1}} \otimes {\Sigma ^t}}>0$), and thus we can conclude that $Q>0$, which implies ${\tilde{D}_K Q  \tilde{D}_K^\top}>0$, because $\tilde{D}_K$ is full row rank.
Then, we need to prove that the Schur complement $\Sigma _K^{\tilde{y},s} / ({\tilde{D}_K Q  \tilde{D}_K^\top})$, is positive definite. This Schur complement is given by
\begin{align}
&\Sigma_K^{\tilde{y},s} / ({\tilde{D}_K Q  \tilde{D}_K^\top}) =
 I_K \otimes \Sigma^w + {{\Sigma }^v_K} +\\ \nonumber
& \tilde{C}_K \left( {Q - Q \tilde{D}_K^\top \left({\tilde{D}_K Q \tilde{D}_K^\top} \right)^{-1}\tilde{D}_K Q} \right) \tilde{C}_K^\top.
\end{align}
Since $(I_K \otimes \Sigma^w) + {{\Sigma }^v_K}$ is positive definite, a sufficient condition for $\Sigma _K^{\tilde{y},s} / ({\tilde{D}_K Q  \tilde{D}_K^\top})>\mathbf{0}$ is
\begin{eqnarray}
Q'' := \left( {Q - Q \tilde{D}_K^\top \left( {\tilde{D}_K Q \tilde{D}_K^\top} \right)^{-1}\tilde{D}_K Q} \right)>\mathbf{0}.
\end{eqnarray}
Regarding ${Q''}$ as the Schur complement of a higher dimensional matrix ${Q'''}$, we can conclude that:
\begin{eqnarray}
\begin{array}{l}
Q'' > \mathbf{0} \Leftrightarrow Q''' = \left[ {\begin{array}{*{20}{c}}
Q&{Q\tilde{D}_K^\top}\\
{\tilde{D}_K Q}&{\tilde{D}_K Q \tilde{D}_K^\top}
\end{array}} \right]\\
 = \left[ {\begin{array}{*{20}{c}}
Q\\
{\tilde{D}_K Q}
\end{array}} \right]{Q^{ - 1}}\left[ {\begin{array}{*{20}{c}}
Q&{Q\tilde{D}_K^\top}
\end{array}} \right] > \mathbf{0},
\end{array}
\end{eqnarray}
which is trivially true because $Q^{-1}$ is positive definite since $Q>\mathbf{0}$. Hence, ${\tilde{D}_K Q \tilde{D}_K^\top}$ and $\Sigma _K^{\tilde{y},s} / ({\tilde{D}_K Q  \tilde{D}_K^\top})$ are both positive definite, and thus $\Sigma _K^{\tilde{y},s}>\mathbf{0}$.
\hfill $\blacksquare$\\
To obtain the densities of $\tilde{y}^K$ and $s^K$ and their cross-covariance $\Sigma^{s\tilde{y}}_K$, we just marginalize (see \cite{Ross}) their joint density in Lemma 1 over $s_K$ and $\tilde{y}_K$, see the following corollary.
\begin{corollary} \label{corollary1}
$\tilde{y}^{K} \sim \mathcal{N} [\mu^{\tilde{y}}_K, \Sigma^{\tilde{y}}_K]$ and
$s^{K} \sim \mathcal{N}[\mu^s_K,\Sigma^s_K]$\emph{:}
\begin{align}
&\mu^{\tilde{y}}_K = \tilde{C}_K F_K \mu^x_1 + \tilde{C}_K N_K u^{K-1},\nonumber\\
&\mu^s_K = \tilde{D}_KF_K \mu^x_1 + \tilde{D}_KN_Ku^{K-1},\nonumber\\
&\Sigma^{\tilde{y}}_K = \begin{pmatrix} I_{n_y}  & \mathbf{0} \end{pmatrix} \Sigma^{\tilde{y},s}_K \begin{pmatrix} I_{n_y} & \mathbf{0}  \end{pmatrix}^\top \in \mathbb{R}^{Kn_y \times Kn_y},\nonumber\\
&\Sigma^s_K = \begin{pmatrix} \mathbf{0}  & I_{n_s} \end{pmatrix} \Sigma^{\tilde{y},s}_K \begin{pmatrix} \mathbf{0} & I_{n_s}  \end{pmatrix}^\top \in \mathbb{R}^{Kn_s \times Kn_s},\nonumber\\
&\Sigma^{s\tilde{y}}_K = \begin{pmatrix}  \mathbf{0} & I_{n_s} \end{pmatrix} \Sigma^{\tilde{y},s}_K \begin{pmatrix}   I_{n_y} & \mathbf{0} \end{pmatrix}^\top \in \mathbb{R}^{Kn_s \times Kn_y}.
\end{align}
\end{corollary}
At this point, we have the three covariance matrices $(\Sigma^{\tilde{y}}_K,\Sigma^s_K,\Sigma^{s\tilde{y}}_K)$ that we need to compute the mutual information $I\left[ {{s^K};{\tilde{y}^K}} \right]$. Note that $\Sigma^s_K$ and $\Sigma _K^{s\tilde{y}}$ are independent of the design variables $\Sigma_K^v$ and $\Sigma_K^j$, and $\Sigma_K^{\tilde{y}}$ depends on $\Sigma_K^{v}$:
\begin{align}
&\Sigma _K^{\tilde{y},s} = \begin{pmatrix} \Sigma _K^{\tilde{y}} & {\Sigma _K^{s\tilde{y}}}^\top \\ \Sigma _K^{s\tilde{y}} & \Sigma _K^{s} \end{pmatrix},\label{SigmaS_Z}\\[1mm]
&\Sigma_K^{\tilde{y}} = {\tilde{C}_K Q \tilde{C}_K^\top +  (I_K \otimes \Sigma^w) + {{\Sigma }^v_K}},\label{eq3g}\\[1mm]
&\Sigma^s_K = { \tilde{D}_K Q  \tilde{D}_K^\top},\label{eq3h}\\[1mm]
&\Sigma _K^{s\tilde{y}} = {  \tilde{D}_K Q  \tilde{C}_K^\top},\label{SigmaSZ}
\end{align}
with $Q$ as in \eqref{Q} independent of $\Sigma^v_K$. Then, given $(\Sigma_K^{\tilde{y},s},\Sigma^{\tilde{y}}_K,\Sigma^s_K,\Sigma^{s\tilde{y}}_K)$ in \eqref{SigmaS_Z}-\eqref{SigmaSZ}, we can write $I[{s^K};{\tilde{y}^K}]$ as follows \cite{Cover}:
\begin{subequations}
\begin{align}
&I[{s^K};{\tilde{y}^K}] = h\left[ s^K \right] + h\left[\tilde{y}^K \right] - h[\tilde{y}^K,s^K]\label{eq3a}\\[1mm]
& = \frac{1}{2} \log{\det \left( {\Sigma _K^s} \right)} - \frac{1}{2} \log{\det  {\left( {\Sigma} _K^s - {{\Sigma} _K^{s\tilde{y}}} {{\Sigma} _K^ {\tilde{y}}}^{-1} {{\Sigma} _K^{s\tilde{y}}}^\top \right),}}\label{eq3f}
\end{align}
\end{subequations}
where \eqref{eq3f} follows from standard determinant and logarithm formulas. In the following lemma, we prove that minimizing the cost function $I[s^K;\tilde{y}^K] - h[j^K]$ using $({{\Sigma }^v_K},{{\Sigma }^j_K})$ as optimization variables is equivalent to solving a convex program subject to Linear Matrix Inequalities (LMI) constraints.
\begin{lemma}\label{mutualinformationcov}
Minimizing $I[s^K;\tilde{y}^K] - h[j^K]$ is equivalent to solving the following convex program:
\begin{eqnarray}
\left\{\begin{aligned}
	&\min_{\Sigma^j_K, {\Pi _K},\Sigma_K^{v}}\
       - \log{\det \left({\Pi _K} \right)}-\log \det \left( \Sigma^j_K \right) \label{finalcost_program}\\[1mm]
    &\hspace{4mm}\text{\emph{s.t. }}  \begin{bmatrix}
\Sigma _K^s - \Pi _K & \Sigma_K^{s\tilde{y}}\\[2mm]
{\Sigma_K^{s\tilde{y}}}^\top & \Sigma_K^{\tilde{y}}
\end{bmatrix} \geq \mathbf{0} ,\Pi _K > \mathbf{0}.
\end{aligned}\right.
\end{eqnarray}
\end{lemma}
\textbf{\emph{Proof}}: {(a)} For any positive definite matrix $\Sigma$, the function $-\log\det(\Sigma)$ is convex in $\Sigma$ \cite{boyd2004convex}. It follows that $-h[j^K]$ in \eqref{hUprim} is a convex function of $\Sigma^j_K$. Minimizing $-h[j^K]$ amounts to minimizing $-\log \det \left( \Sigma^j_K \right)$ as all other terms in \eqref{hUprim} are constants. \\
{(b)} Next, consider the expression for $I[\tilde{y}^K;s^K]$ in \eqref{eq3f}. Due to the monotonicity of the determinant function and the fact that $\Sigma _K^s$ is independent of the design variables, minimizing \eqref{eq3f} is equivalent to
\begin{eqnarray}
\left\{\begin{aligned}
	&\min_{{\Pi _K},\Sigma_K^{v}}\
      - \log{\det \left({\Pi _K} \right)} \label{epigraphcost}\\[1mm]
    &\hspace{4mm}\text{s.t. } \mathbf{0} < {\Pi _K} \le  {\left( {\Sigma} _K^s - {{\Sigma} _K^{s\tilde{y}}} {{\Sigma} _K^ {\tilde{y}}}^{-1} {{\Sigma} _K^{s\tilde{y}}}^\top \right).} \label{inequalityofcost}
\end{aligned}\right.
\end{eqnarray}
{(c)} The inequality term in \eqref{inequalityofcost} can be rewritten using Schur complement properties \cite{zhang2006schur} as
\begin{eqnarray}
&\left[ {\begin{array}{*{20}{c}}
{\Sigma _K^s} - {\Pi _K} & {{\Sigma} _K^{s\tilde{y}}}\\
{{\Sigma} _K^{s\tilde{y}}}^\top&{{\Sigma} _K^ {\tilde{y}}}
\end{array}} \right] \ge \mathbf{0} \, {,{\Pi _K} > \mathbf{0}}. \label{finalcost}
\end{eqnarray}
Since $\Sigma_K^{s\tilde{y}}$ is constant in the design variables (see \eqref{SigmaSZ}), and $\Sigma_K^{\tilde{y}}$ is convex in $\Sigma_K^{v}$, we can conclude that minimizing $I[s^K;\tilde{y}^K] - h[j^K]$ is equivalent to solving the convex program in \eqref{finalcost_program}. \hfill $\blacksquare$

By Lemma \ref{mutualinformationcov}, minimizing the cost in \eqref{optimizationproblem} is equivalent to solving the convex program in \eqref{finalcost_program}. Then, if the distortion constraint in \eqref{optimizationproblem} is convex in the design variables, we can find optimal distorting mechanisms efficiently using off-the-shelf optimization algorithms.
\subsection{Detection Performance Constraints: Formulation and Convexity}
We consider a constraint on the false alarm rate in each time step, see \eqref{optimizationproblem}, as a metric of performance degradation of the detector. This constraint is fully characterized by the CDF of the distorted distance measure $\tilde{z}_k$. Unfortunately, no closed-form expression is available for this CDF. To address this, we seek a lower bound on the CDF of $\tilde{z}_k$ to use it to enforce constraint \eqref{constraint}. Define a standard Gaussian vector $m_k \sim \mathcal{N}(\mathbf{0},I_{n_y})$.
Since $\tilde{r}_k\sim \mathcal{N}(\mathbf{0},\tilde{\Sigma}_k)$, it is easy to verify that $\tilde{r}_k$ can be written in terms of $m_k$ as follows:
\begin{equation}
    \tilde{r}_k=\left(\tilde{\Sigma}_k\right)^{1/2} m_k,
\end{equation}
where $\left(\tilde{\Sigma}_k\right)^{1/2}$ is the square root of $\tilde{\Sigma}_k$. Hence, $\tilde{z}_k$ can be written as:
\begin{align}
     \tilde{z}_{k} &= \tilde{r}^{\top}_{k} \Sigma^{-1} \tilde{r}_{k}=m_k^\top (\tilde{\Sigma}^{1/2}_k)^\top \Sigma^{-1} \tilde{\Sigma}^{1/2}_k m_k.
\end{align}
Define the sequence $q_k := \beta m_k^\top m_k$, $\beta \in \mathbb{R}^+$, $k \in \mathcal{K}$, and let $\Sigma'_k := \left((\tilde{\Sigma}_k)^{1/2}\right)^\top \Sigma^{-1} (\tilde{\Sigma}_k)^{1/2}$; then, $\tilde{z}_{k} \le q_k$ for all $k \in \mathcal{K}$ implies $m_k^\top \Sigma'_k m_k \le \beta m_k^\top m_k$.
It follows that $\tilde{z}_{k} \le q_k$ if and only if
\begin{eqnarray}
     \Sigma'_k \le \beta {I}_{n_y}.\label{eq:Mbeta}
\end{eqnarray}
To be able to find the constraints of Problem $1$ in terms of the CDF of the upper bound $q_k$, we need to find a relation between the CDFs of $\tilde{z}_{k}$ and $q_k$.
In the next lemma, we show such a relation.
\begin{lemma}\label{CDFinequality}
$\tilde{z}_{k} \le q_k$, $k \in \mathcal{K}$, implies\emph{:}
\begin{equation}
 F_{\tilde{z}_{k}} (\alpha) \ge F_{q_k} (\alpha), \label{eq:cdf}
\end{equation}
where $F_{\tilde{z}_{k}} (\alpha)$ and $F_{q_k}(\alpha)$ denote the CDFs of $\tilde{z}_{k}$ and $q_k$, respectively.
\end{lemma}
\emph{\textbf{Proof}}: The proof is given in Appendix A.\hfill $\blacksquare$

From \eqref{eq:Mbeta}, Lemma \ref{CDFinequality}, and \eqref{constraint}, we can conclude that the constraints of the optimization problem \eqref{optimizationproblem} hold if $\Sigma'_k \le \beta {I}_{n_y}$ and
\begin{equation}
    F_{\beta m_k^\top m_k} (\alpha) > 1 - \mathcal{A}^*   - \epsilon \,\,\text{for all }\,k \in \mathcal{K}.\label{eq:beta}
\end{equation}
Therefore, if we choose $\beta$ to satisfy \eqref{eq:beta}, the false alarm constraint in \eqref{optimizationproblem} translate to $\Sigma'_k \le \beta {I}_{n_y}$. Given a desired distortion level $\epsilon$ and desired false alarm rate $\mathcal{A}^*$, a lower bound on $F_{\beta m_k^\top m_k} (\alpha)$ can be computed. Since $m_k^\top m_k$ follows a chi-squared distribution with $n_y$ degree of freedom, 
the CDF $F_{q_k} (\alpha)$ can be computed as follows (see \cite{stacy1962generalization}):
\begin{equation}\label{eq:findbeta}
  F_{q_k} (\alpha)=P\left({n_y}/{2},{\alpha}/{2\beta}\right),
\end{equation}
where $P(\cdot,\cdot)$ is the regularized gamma function \cite{stacy1962generalization}. Substituting \eqref{eq:findbeta} into \eqref{eq:beta} yields:
\begin{equation}\label{largestbeta}
   P\left({n_y}/{2},{\alpha}/{2\beta}\right) > 1 - \mathcal{A}^*   - \epsilon.
\end{equation}
Because the regularized gamma function is an increasing function of $\frac{\alpha}{2\beta}$ \cite{furman2008monotonicity}, the largest $\beta$ satisfying \eqref{largestbeta} for given $\alpha$, $\epsilon$, and $\mathcal{A}^*$ can be calculated using the inverse of the lower incomplete gamma function \cite{furman2008monotonicity} as follows:
\begin{equation}\label{eq:betaf}
    \beta < \beta^* := \frac{\alpha}{2 P^{-1} \left(\frac{n_y}{2},1 - \mathcal{A}^*   - \epsilon\right)}.
\end{equation}
Then, once we have $\beta^*$, the false alarm constraint in Problem $1$ amounts to $\Sigma'_k \le \beta {I}_{n_y} < \beta^* {I}_{n_y}$,   
which is a characterization in terms of the design variables (as $\Sigma'_k$ depends on $\Sigma^v_k$ and $\Sigma^j_k$). We state this explicitly in the following lemma.
\begin{lemma}\label{constraints}
Inequality $\Sigma'_k=(\tilde{\Sigma}^{1/2}_k)^\top \Sigma^{-1} \tilde{\Sigma}^{1/2}_k < \beta^* {I}_{n_y}$ can be written in terms of $\Sigma^v_k$ and $\Sigma^j_k$ as:
\begin{equation}
    \tilde{\Sigma}_k < \beta^* \Sigma, \label{eq:lemma41}
\end{equation}
\end{lemma}
with $\beta^*$ defined in \eqref{eq:betaf} and $\tilde{\Sigma}_k=\Sigma + \Sigma^v_k + C L\Sigma^v_k L^\top C^\top + CB\Sigma^j_k B^\top C^\top$.\\
\emph{\textbf{Proof}}: The proof is given in Appendix B.\hfill $\blacksquare$\\

By Lemma \ref{mutualinformationcov} and Lemma \ref{constraints} the cost $I[s^K;\tilde{y}^K] - h[j^K]$ and false alarm constraint $F_{\tilde{z}_k}(\alpha)> 1 - \mathcal{A}^* - \epsilon$ are written linearly in terms of the optimization variables ${{\Sigma }^v_K}$ and ${{\Sigma }^j_K}$.
In what follows, we pose the complete nonlinear convex program to solve Problem $1$.
\begin{theorem}\label{th3}
Consider the system dynamics \eqref{eq1}, distorting mechanism \eqref{eq2}, $k \in \mathcal{K}$ with $\mathcal{K}=\{1,\ldots,K\}$ and $K\in \mathbb{N}$, desired false alarm rate $\mathcal{A}^*$, maximum distortion level ${\epsilon} \in {\mathbb{R}^+}$, the chi-squared procedure threshold $\alpha$, the anomaly detector Kalman filter \eqref{eq5}, and covariance matrices $\Sigma^{\tilde{y}}_K$, $\Sigma _K^{s}$, and $\Sigma _K^{s\tilde{y}}$ \emph{(}given in \eqref{SigmaS_Z}-\eqref{SigmaSZ}\emph{)}. Then, the covariance matrices ${{\Sigma }^v_K}>0$ and ${{\Sigma }^j_K}>0$ that minimize $I[s^K;\tilde{y}^K] - h[j^K]$ subject to $F_{\tilde{z}_k}(\alpha)> 1 - \mathcal{A}^*   - \epsilon$ can be found by solving the convex program in \eqref{eq:convex_optimization15}.
\end{theorem}
\emph{\textbf{Proof:}} The expressions for the cost and constraints and their convexity follow from Lemma \ref{mutualinformationcov} and Lemma \ref{constraints}.  \hfill $\blacksquare$

\begin{table}
\noindent\rule{\hsize}{1pt}
\begin{equation} 
\begin{aligned}
	&\min_{{\Pi _K}, \Sigma_K^{v}, {{\Sigma }^j_K}}  -\log\det [ \Sigma^j_k ] - \log\det [\Pi_K ],\\[1mm]
    &\text{ s.t. }\left\{\begin{aligned}
    &{\Pi _K} > \mathbf{0},\\ &\begin{bmatrix}
\Sigma _K^s - \Pi _K & \Sigma_K^{s\tilde{y}}\\
{\Sigma_K^{s\tilde{y}}}^\top & \Sigma_K^{\tilde{y}}
\end{bmatrix} \geq \mathbf{0}, \\[1mm]
&\Sigma _K^v > \mathbf{0}, \\[1mm]
&\Sigma _K^j > \mathbf{0}, \\[1mm]
&\Sigma + \Sigma^v_k + C L\Sigma^v_k L^\top C^\top + CB\Sigma^j_k B^\top C^\top < \beta^* \Sigma, \\[1mm]
&\beta^*=\frac{\alpha}{2 P^{-1} \left(\frac{n_y}{2},1 - \mathcal{A}^*   - \epsilon\right)}.\end{aligned}\right.
\end{aligned}\label{eq:convex_optimization15}
\end{equation}
\noindent\rule{\hsize}{1pt}
\end{table}
Solving the convex program in \eqref{eq:convex_optimization15} provides an upper bound on the solution for the original optimization problem in \eqref{optimizationproblem}, considering the relaxation of constraints. This approach allows us to navigate the challenges of solving the original optimization problem, which is nonconvex in cost and constraints. Therefore, Theorem $1$ provides a tool to find suboptimal privacy-preserving mechanisms to achieve a trade-off between privacy and anomaly detector performance. In the following, we significantly generalize the class of stochastic dynamical systems that Theorem $1$ can deal with. 
\begin{remark}
In the following proposition, we prove that we can employ Theorem $1$ to find optimal privacy-preserving mechanisms for prior distributions that are not necessarily Gaussian, as long as they are log-concave \emph{\cite{prekopa1980logarithmic}}. In particular, we prove that if $s^K$ and $\tilde{y}^K$ follow non-Gaussian log-concave distributions, their mutual information is upper-bounded by an affine function of the mutual information between
Gaussian random vectors with the same mean and covariance matrices as $s^K$ and $\tilde{y}^K$, respectively.
Examples of log-concave distributions are normal, exponential, uniform, Laplace, chi, beta, and logistic distributions. We refer the reader to \emph{\cite{LogConcave,wellner2012log}} for a comprehensive list and properties of log-concave distributions.
\end{remark}
\begin{proposition}\label{mutualinformationupper}
Let $s^K$ and $\tilde{y}^K$ be jointly distributed following log-concave multivariate probability distribution. The mutual information $I[s^K;\tilde{y}^K]$ can be upper-bounded as follows:
\begin{eqnarray}
   I[s^K;\tilde{y}^K] \le I[s^K_{\mathcal{N}};\tilde{y}^K_{\mathcal{N}}] +  C_n,\label{eqmutualinfupper1}
\end{eqnarray}
where $C_n = \frac{1}{2}\log(2\pi e c(n))^n$, $n=K(n_s + n_y)$, $c(n)=\frac{e^{2} n^{2}}{4 \sqrt{2}(n+2)}$, and $s^K_{\mathcal{N}}$ and $\tilde{y}^K_{\mathcal{N}}$ denote Gaussian random vectors with the same covariance matrices as $s^K$ and $\tilde{y}^K$.
\end{proposition}
\textbf{\emph{Proof}}: The proof is given in Appendix C.
\hfill $\blacksquare$

\begin{remark}\label{upperboundremark1}
Proposition 1 implies that by minimizing $I[s^K_{\mathcal{N}};\tilde{y}^K_{\mathcal{N}}]-h[j^K]$ (based on the optimization problem in Theorem 1), we can effectively decrease the information leakage $I[s^K;\tilde{y}^K]-h[j^K]$ for any $(s^K,\tilde{y}^K)$ following log-concave distributions.
\end{remark}

A tighter bound for the mutual information of log-concave random vectors can be achieved based on their maximum density. The maximum density of a random vector is calculated by its $L_\infty$ norm \cite{bobkov2011entropy}. In the following proposition, we prove that the mutual information between log-concave random vectors $s^K$ and $\tilde{y}^K$ can be upper-bounded by an affine function of the mutual information of Gaussian random vectors with maximum density being the same as those of $s^K$ and $\tilde{y}^K$.
\begin{proposition}\label{mutualinformationupper2}
Let $s^K$ and $\tilde{y}^K$ be jointly distributed following a log-concave multivariate joint probability distribution with mutual information $I[s^K;\tilde{y}^K]$, then:
\begin{eqnarray}
   I[s^K;\tilde{y}^K] \le  I[s^K_{{\mathcal{N}}^*};\tilde{y}^K_{{\mathcal{N}}^*}] + n,\label{eqmutualinfupper}
\end{eqnarray}
where $n=K(n_s + n_y)$, and $s^K_{{\mathcal{N}}^*}$ and $\tilde{y}^K_{{\mathcal{N}}^*}$ denote multivariate normally distributed random vectors with maximum density being the same as those of $s^K$ and $\tilde{y}^K$, respectively.
\end{proposition}
\textbf{\emph{Proof}}: The proof is given in Appendix D. \hfill $\blacksquare$
\begin{remark}\label{upperboundremark}
In Proposition 2 we prove that the difference between $I[s^K;\tilde{y}^K]$ and $I[s^K_{{\mathcal{N}}^*};\tilde{y}^K_{{\mathcal{N}}^*}]$ can be upper-bounded by $n$. This upper bound is still conservative since it needs to cover all log-concave distributed random vectors. However, the authors in \cite{bobkov2011entropy} observe that every log-concave random vector has approximately the same entropy per coordinate as a related Gaussian vector (which resulted in approximately the same mutual information), and log-concave distributions resemble Gaussian distributions. Therefore, by minimizing $I[s^K_{{\mathcal{N}}^*};\tilde{y}^K_{{\mathcal{N}}^*}]$ (based on the optimization problem in Theorem 1), we can effectively decrease the information leakage $I[s^K;\tilde{y}^K]-h[j^K]$ for any $(s^K,\tilde{y}^K)$ following a log-concave distribution.
\end{remark}
\begin{remark}\label{arbitrarylargetimehorizonremark}
In this manuscript, we have addressed the problem of designing optimal probabilistic mappings to maximize privacy for a finite time horizon. In \cite{hayati2022privacy}, it has been proven that the proposed mechanisms can be implemented for an arbitrarily large time horizon by solving finite horizon problems repeatedly and sequentially implementing the obtained mechanisms.
\end{remark}
\begin{remark}
In this work, we consider the problem of solving an optimization problem to hold a trade-off between privacy and the performance of the anomaly detection algorithm in order to design optimal privacy-distorting mechanisms. For designing the privacy mechanisms in the general case, i.e., for utilities of data other than anomaly detection, another optimization problem to hold a trade-off between \emph{privacy} and the amount of \emph{distortion} is described in \cite{hayati2021finite,hayati2022gaussian}. This generalizes the applicability of the results in the paper to a much wider range of applications (beyond anomaly detection) in which a balance between privacy and performance (challenged by the distortion of data) is needed.
\end{remark}
\section{On Anomaly Detection Performance Metrics}
In order to assess the cost of the proposed privacy-preserving mechanisms in terms of detectability loss, three different metrics are employed.

The first metric for evaluating the performance of the anomaly detector is the false alarm rate. In the absence of privacy mechanisms, the false alarm rate is $\mathcal{A}^*$. After applying the privacy mechanisms, based on distortion constraint $\operatorname{Pr}[\tilde{z}_k  > \alpha]<\mathcal{A}^* + \epsilon$ in optimization problem \eqref{optimizationproblem}, the false alarm rate $\mathcal{A}=\operatorname{Pr}[\tilde{z}_k > \alpha]$ can be increased by $\epsilon$. Therefore, the privacy distorting mechanism degrades the performance of the anomaly detector in terms of the false alarm rate. We calculate the false alarm rate based on the CDF of $\tilde{z}_k$ as follows:
\begin{align}\label{falsealarmwprivacywfault}
  \mathcal{A}=\operatorname{Pr}\left[\tilde{z}_k > \alpha|\delta_k=\mathbf{0}\right]=1-F_{\tilde{z}_k}\left(\alpha|\delta_k=\mathbf{0}\right).  
\end{align}
Since $\tilde{z}_k$ approximately follows a gamma distribution (see Appendix E), its CDF can be determined by the inverse of the lower incomplete gamma function \cite{stacy1962generalization}.

Another metric is the detection rate of the anomaly detector for deterministic additive fault $\delta_k$. The detection rate to detect faulty systems with distance measure ${z}_{k}$ is defined as $\mathcal{B}=\operatorname{Pr}\left[{z}_{k} > \alpha|\delta_k \neq \mathbf{0} \right]$. 
We start with the case when there is no privacy distortion in the system but $\delta_k \neq \mathbf{0}$. In this case, the faulty residual sequence is determined by \eqref{eq7}. According to the superposition principle, the faulty residual sequence can be written as:
\begin{align} \label{eq:superpositionrk}
    r_k=r_k^\delta +r_k^n,
\end{align}
where $r_k^\delta$ is a signal representing the change in the residual due to the fault $\delta_k$, and $r_k^n$ represents the change in the residual due to system and measurement noises. In \eqref{eq9}, we proved that if there is no anomaly and no privacy distortion in the system, the residual follows Gaussian distribution ${r}_k \sim \mathcal{N}(\mathbf{0},\Sigma)$. Besides, since $\delta_k$ is a deterministic additive fault, $r_k^\delta$ will only appear in the mean of the residual. Therefore, the residual follows Gaussian distribution ${r}_k \sim \mathcal{N}(r_k^\delta,\Sigma)$ when $\delta_k \neq \mathbf{0}$. It can be proved that based on this new residual, the new distance measure is distributed as a non-central chi-squared with $n_y$ degrees of freedom and non-centrality parameter $\lambda_{k} =\left\|\Sigma^{(-1 / 2)} r_k^\delta\right\|^{2}$ \cite{feiveson1968distribution}.
Note that for a fixed $\Sigma$, $\lambda_{k}$ is an increasing function of the fault. 
Let $\mathcal{B}^*$ be the detection rate without privacy distortion. This detection rate can be calculated based on the CDF of $z_k$ as
\begin{align}\label{detectionratewoprivacy}
   \mathcal{B}^*=1-F_{{z}_{k}}\left(\alpha|\delta_k \neq \mathbf{0}\right). 
\end{align}
Since ${z}_{k}$ is a non-central chi-squared variable for $\delta_k \neq \mathbf{0}$, and due to the fact that the CDF of a non-central chi-squared variable is a decreasing function of its non-centrality parameter (see \cite{samir1990computation}), the CDF of ${z}_{k}$, $F_{{z}_{k}}\left(\alpha|\delta_k \neq \mathbf{0}\right)$, is a decreasing function of $\lambda_{k}$. As a result, by increasing $\lambda_{k}$, $\mathcal{B}^*=1-F_{{z}_{k}}\left(\alpha|\delta_k \neq \mathbf{0}\right)$ increases. Therefore, given the fact that $\mathcal{B}^*$ is an increasing function of $\lambda_{k}$ and $\lambda_{k}$ is an increasing function of $\delta_k$, we can conclude that for larger fault $\delta_k$, detection rate $\mathcal{B}^*$ will be increased, which amounts to better detectability.\\
For the case when there is both privacy distortion and fault in the system, based on the superposition principle, we need to consider another term in the residual for the additive noises for privacy compared with \eqref{eq:superpositionrk}. Therefore, the distorted faulty residual is given by:
\begin{align}
    \tilde{r}_k=r_k^\delta +r_k^n + r_k^p,
\end{align}
where $r_k^p$ is a signal representing the change in the residual due to privacy distortions. In Equation \eqref{eq12}, we showed that the distorted residual follows $\tilde{r}_k \sim \mathcal{N}(\mathbf{0},\tilde{\Sigma}_k)$. Consequently, the distorted faulty residual follows Gaussian distribution as $\tilde{r}_k \sim \mathcal{N}(r_k^\delta,\tilde{\Sigma}_k)$. It can be proved that based on the new residual, the new distance measure $\tilde{z}_{k} ={\tilde{r}} ^{\top}_{k} \Sigma^{-1} \tilde{r}_{k}$ is distributed as a generalized chi-squared \cite{imhof1961computing}.
Then, the detection rate $\mathcal{B}$ in the presence of privacy mechanisms can be calculated as follows:
\begin{align}\label{detectionratewprivacywfault}
    \mathcal{B}=1-F_{\tilde{z}_k}\left(\alpha|\delta_k \neq \mathbf{0}\right).
\end{align}
Since $\tilde{z}_k$ follows generalized chi-squared, its PDF and CDF, unfortunately, cannot be expressed in closed form, although approximations are available \cite{imhof1961computing}. 

To determine the effect of privacy distortion on the detection rate, the function of the detection rate in terms of $\delta_k$ in with and without privacy distortion cases can be compared. At the start point, when $\delta_k=\mathbf{0}$, the detection rate equals the false alarm rate, and therefore, the detection rate is higher when privacy distortions are added. Besides, as we mentioned before, when there is no privacy distortion, the detection rate is monotonically increasing for an increasing amount of $\delta_k$. In addition, when $\delta_k$ goes to infinity, the detection rate goes to $1$ for both with and without privacy distortions cases. Hence, by adding privacy distortions, the function of detection rate in terms of fault starts from a higher point (compared to without privacy distortions case) and reaches the same point for infinitely large fault where $\mathcal{B}=1$. Hence, the privacy distortion flattens the detection rate function in terms of fault magnitude (see Fig. \ref{detectionrate2} for our implemented numerical example of this occurrence). The detection rate could not be employed as a performance metric in the privacy-utility trade-off optimization problem. Because to be able to use it, we need to know the distribution of the fault, which is not a realistic assumption. Moreover, even when the fault is deterministic, the detection rate in the presence of privacy mechanisms \eqref{detectionratewprivacywfault} needs to be calculated numerically. Therefore, this metric cannot effectively be employed to define the detection performance constraint as we do with the false alarm rate.

To investigate the overall effect of privacy mechanisms on detection performances, the Receiver Operating Characteristic (ROC) is considered. 
The ROC curves show how the detection rate changes as the false-alarm rate increases. In \eqref{falsealarmwprivacywfault} and \eqref{detectionratewprivacywfault}, the false alarm rate and the detection rate are written in terms of $\alpha$, where $\alpha$ is selected to satisfy the desired false alarm rate $\mathcal{A}^*$. Therefore, calculating the false alarm rate and detection rate for different $\mathcal{A}^*$ will allow characterizing the ROC of the proposed detection scheme. The ROC plot can be employed to explore the net effect of privacy mechanisms on detection performances in terms of false alarm rate and detection rate.
\section{Illustrative case study}
The authors of \cite{chen2012robust,IET_CARLOS_JUSTIN} studied the anomaly detection problem for a well-stirred chemical reactor with a heat exchanger. We use this case study to demonstrate our results. The state, input, and output vectors of the considered reactor are:%
\begin{align*}
\left\{
\begin{array}{ll}
x_k =  \begin{pmatrix} C_0\\T_0\\T_w\\T_m \end{pmatrix},
u_k =   C_u,
y_k =   \begin{pmatrix} C_0\\T_0 \end{pmatrix},
\end{array}
\right.
\end{align*}
where
\begin{equation*}
\left\{
\begin{array}{ll}
\begin{aligned}
&C_0: \text{Concentration of the chemical product},\\
&T_0: \text{Temperature of the product},\\
&T_w: \text{Temperature of the jacket water of the heat exchanger},\\
&T_m: \text{Coolant temperature},\\
&C_u: \text{Inlet concentration of the reactant}.
\end{aligned}
\end{array}
\right.
\end{equation*}
\begin{figure}[!htb]
\centering
  \includegraphics[width=3.2in]{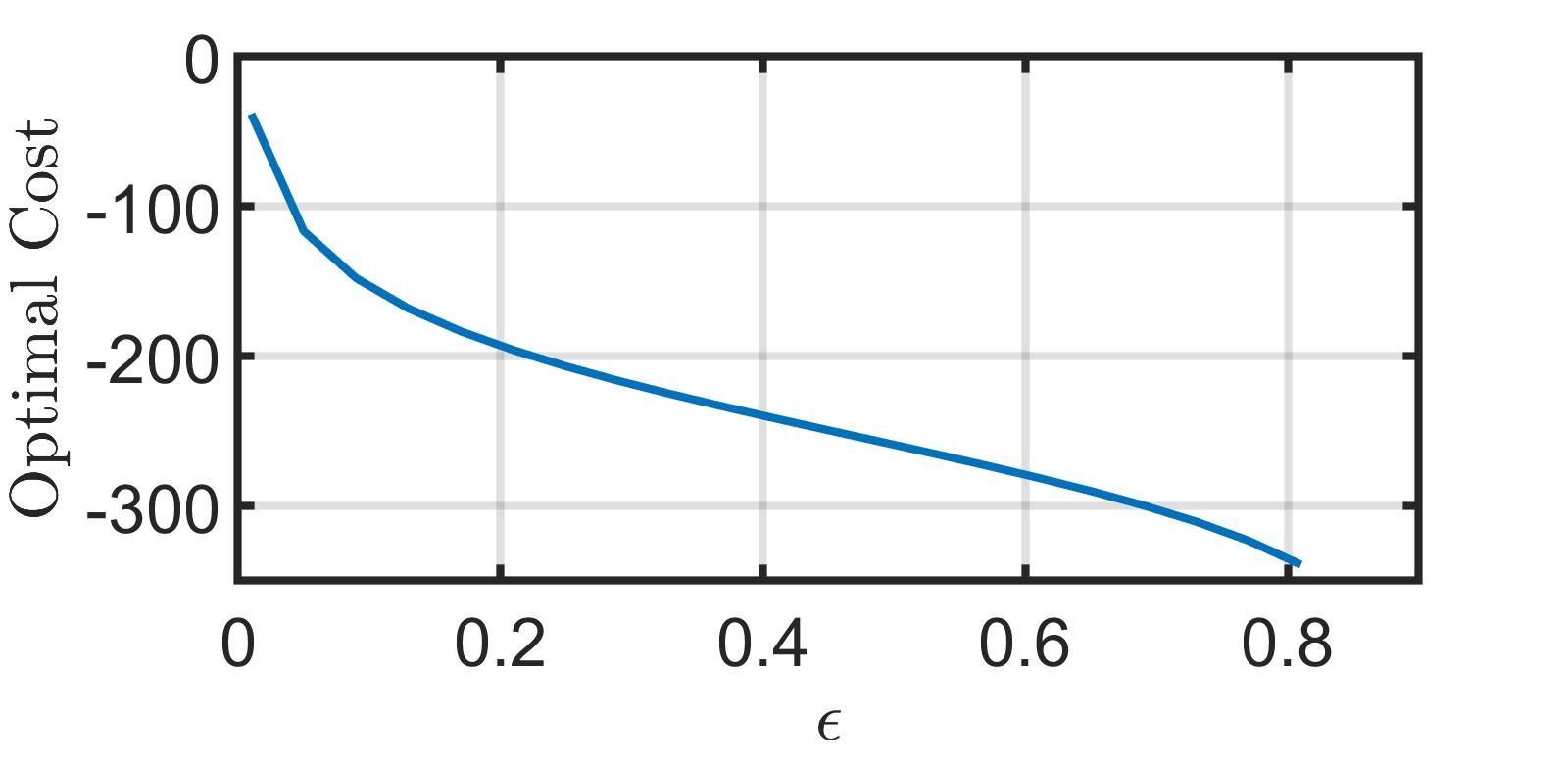}
  \caption{Evolution of the optimal cost function (information leakage) based on increasing $\epsilon$ for horizon $K=60$.}\label{CostEps}
\end{figure}\\
\begin{figure}[!htb]
\centering
  \includegraphics[width=2.8in]{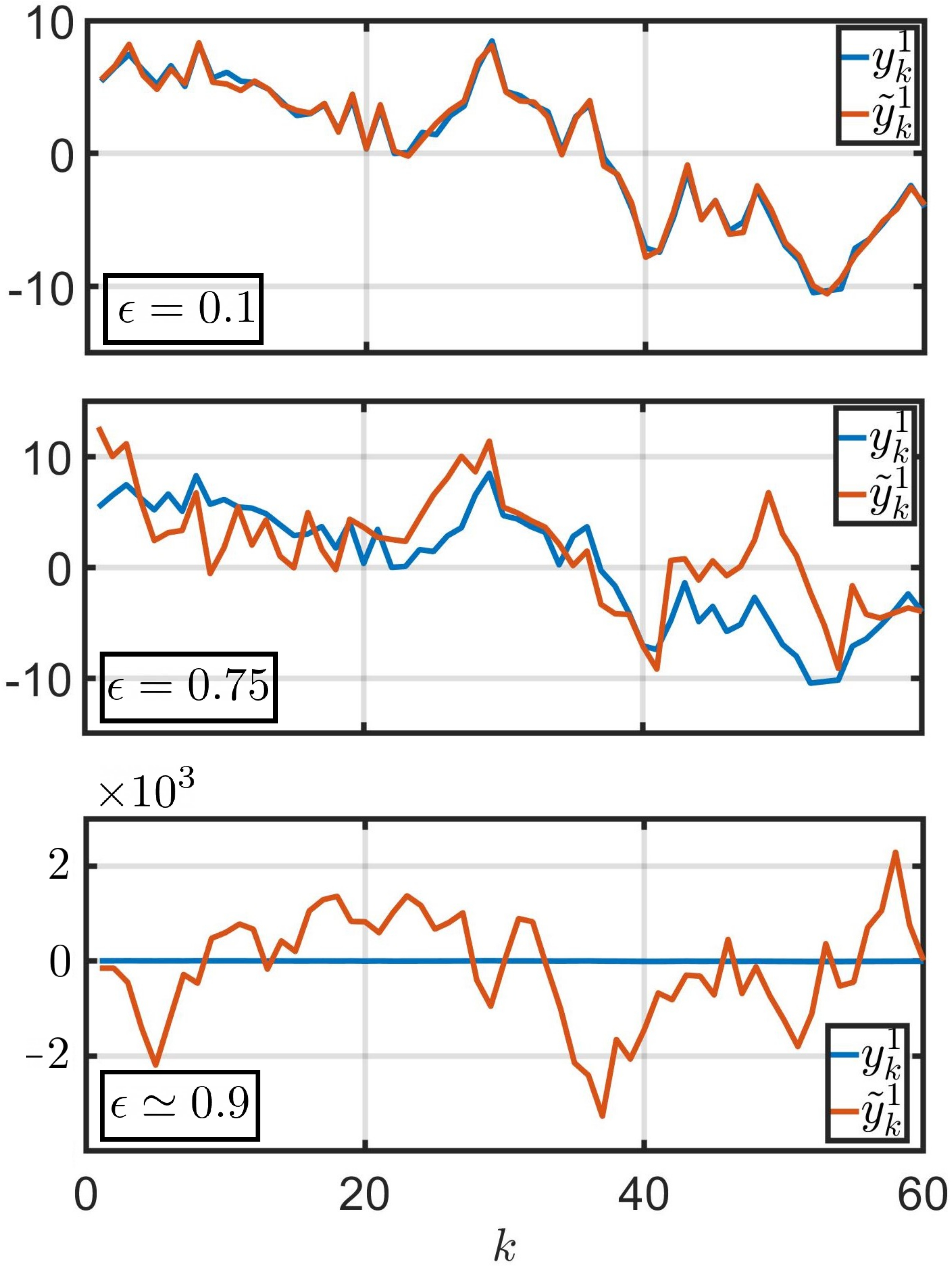}
u  \caption{Comparison between the measurement $y^1_k$ and the distorted measurement $\tilde{y}^1_k$ for different distortion levels.}
\label{YZ}
\end{figure}
\begin{table*}[!ht]
\noindent\rule{\hsize}{1pt}
\begin{equation} \label{eq:experiment}
\left\{\begin{array}{l}
A=\left(\begin{array}{cccc}
0.8353 & 0 & 0 & 0 \\
0 & 0.8324 & 0 & 0.0031 \\
0 & 0.0001 & 0.1633 & 0 \\
0 & 0.0280 & 0.0172 & 0.9320
\end{array}\right), \quad B=\left(\begin{array}{ccc}
0.0458  \\
0  \\
0  \\
0
\end{array}\right), \quad C=\left(\begin{array}{cccc}
1 & 0 & 0 & 0 \\
0 & 1 & 0 & 0
\end{array}\right), \quad \Sigma^w = 0.01I_{n_y},\\
\Sigma^t=\left(\begin{array}{cccc}
0.1 & 0 & 0 & 0 \\
0 & 0.2 & 0 & 0 \\
0 & 0 & 0.3 & 0 \\
0 & 0& 0 & 0.4
\end{array}\right),\quad L =\left(\begin{array}{ccc}
0.8271 & 0  \\
0 & 0.8243  \\
0 & 0.0002  \\
0 & 0.0481
\end{array}\right),  \quad \Sigma=\left(\begin{array}{ccc}
1.0169 & 0 \\
0 & 1.0169
\end{array}\right),\\
x_1 \sim \mathcal{N}[(6.94;13.76;1;1)^\top,I_4], \quad u_k = 50\cos[0.5k]^2, \quad D = (1,0,0,0).
\end{array}\right.
\end{equation}
\noindent\rule{\hsize}{1pt}
\end{table*}
We use the discrete-time dynamics of the reactor introduced in \cite{murguia2021privacy} for this illustrative simulation study with matrices as given in \eqref{eq:experiment}, normally distributed initial condition $x_1 \sim \mathcal{N}[(6.94;13.76;1;1)^\top,I_4]$, and reference signal $u_k = 50\cos[0.5k]^2$. As private output, we use the concentration of the chemical product; then, matrix $D$ in \eqref{eq1} is given by the full row rank matrix $D = (1,0,0,0)$. Therefore, the aim of the privacy scheme is to hide the private state $C_0$, which is the concentration of the reactant, as much as possible without distorting anomaly detection performance excessively.

First, in Figure \ref{CostEps}, we show the effect of false alarm distortion level $\epsilon$ on the amount of optimal information leakage, which is shown by the optimal cost $I[s^K;\tilde{y}^K] - h[j^K]$. This figure depicts the evolution of the optimal cost for increasing $\epsilon$ with time horizon $K=60$ and desired false alarm rate $\mathcal{A}^* = 0.1$ in the absence of fault. As expected, by increasing the distortion level $\epsilon$, which increases the amount of distortion on the false alarm rate as we discussed in Section \Romannum{6} (i.e., leading to poorer anomaly detection performance), the information leakage decreases monotonically (i.e., leading to improved privacy). This figure illustrates that while designing the privacy distorting mechanisms, we need to balance the trade-off between privacy and anomaly detector performance.

The effect of optimal distorting mechanisms is shown in Figure \ref{YZ}, where we contrast actual and distorted measurement data for different distortion levels $\epsilon \in\{0.1,0.75,0.9\}$. We consider $\mathcal{A}^* = 0.1$; therefore, $\epsilon=0.9$ means that the optimization problem in \eqref{eq:convex_optimization15} is solved without considering the distortion constraint (because in this case, the upper bound of the false alarm rate is $\mathcal{A}^*+\epsilon = 1$ which means that there is no constraint on the false alarm rate). As can be seen in these figures, the difference between distorted and actual measurements will increase by increasing the distortion level. The same result is achieved for the comparison of actual and distorted inputs.

\begin{figure}[!t]\centering
  \includegraphics[width=2.9in]{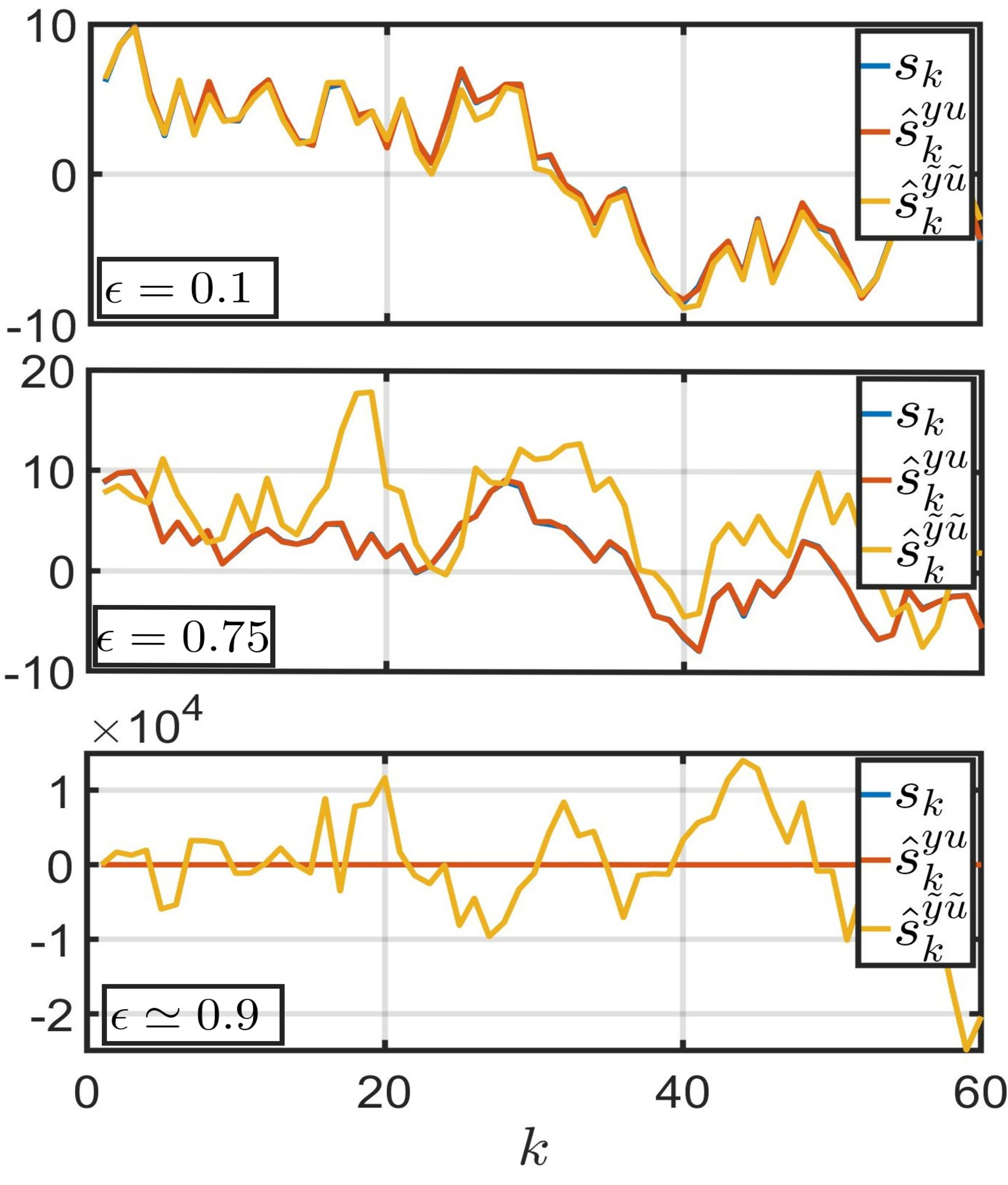}
  \caption{Comparison between the private output $s_k$, its estimate without privacy distortion, ${{\hat s}_k^{yu}}$, and its estimate given distorted data, ${{\hat s}_k^{\tilde{y}\tilde{u}}}$, for different distortion levels $\epsilon$.}\label{estimatedS}
\end{figure}

Next, in Figure \ref{estimatedS}, we depict the realization of $s_k$, the estimated private output without distorting $y_k$ and $u_k$ using the MMSE estimator, ${{\hat s}_k^{yu}}$, and the estimated private output given the distorted vectors $\tilde{y}_k$ and ${\tilde{u}}_k$, ${{\hat s}_k^{\tilde{y}\tilde{u}}}$. We consider time horizon $K=60$, distortion levels $\epsilon \in \{0.1, 0.75,0.9\}$, and desired false alarm rate $\mathcal{A}^*=0.1$ in the absence of fault. As can be seen in these figures, by increasing the distortion level $\epsilon$, the accuracy of estimation decreases. This figure demonstrates that by randomizing measurement and input data, we can prevent worst-case adversaries from accurately estimating $s_k$.

\begin{figure}[!htb]
\centering
  \includegraphics[width=3.2in]{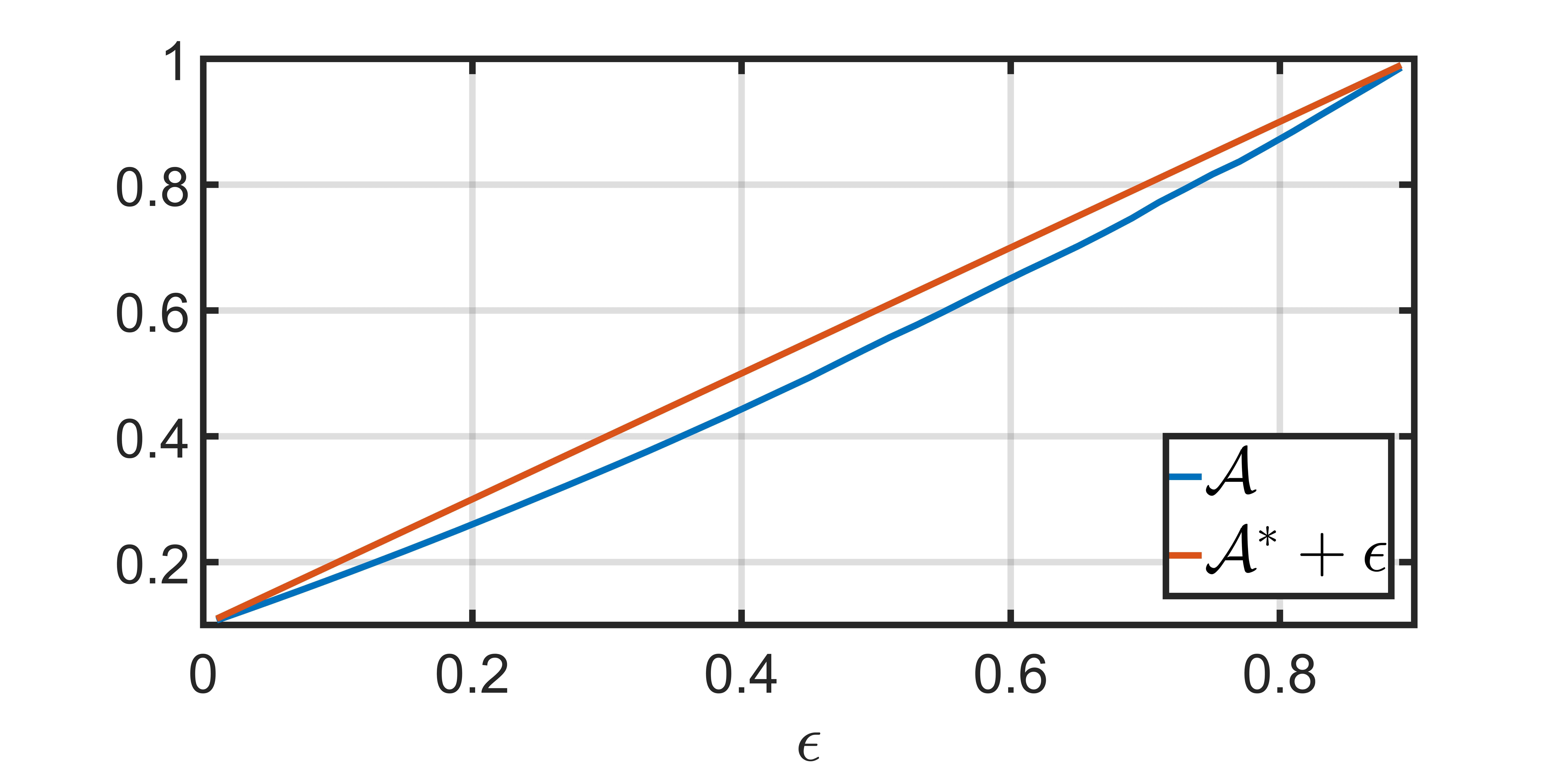}
  \caption{Comparison between expected false alarm rate $\mathcal{A}^*+\epsilon$ and approximation of false alarm rate after privacy distortions $\mathcal{A}$ for an increasing amount of $\epsilon$, horizon $K=60$, and desired false alarm rate $\mathcal{A}^* = 0.1$.}\label{fig:approxfalsealarmrate}
\end{figure}

To evaluate the effect of privacy mechanisms on anomaly detection performance, we apply three different metrics, as discussed in Section \Romannum{6}.
First, in Figure \ref{fig:approxfalsealarmrate}, the effect of distortion level $\epsilon$ on the false alarm rate is shown. As discussed in Section \Romannum{6}, we employ the CDF of the approximation of $\tilde{z}_k$, $F_{\tilde{z}_k}(\alpha)$, using the lower incomplete gamma function to determine the false alarm rate.
As can be seen, by increasing $\epsilon$, the false alarm rate increases. Therefore as we expected, increasing false alarm distortion level $\epsilon$, which resulted in increasing privacy distortion and decreasing the information leakage, indeed reduces the performance of the anomaly detector in terms of false alarm rate. Besides, this figure shows that the upper bound we designed for the false alarm rate in distortion constraint \eqref{constraint}, $\mathcal{A}^* + \epsilon$, is respected and not so conservative.

\begin{figure}[!htb]
  \centering
  \includegraphics[width=3.2in]{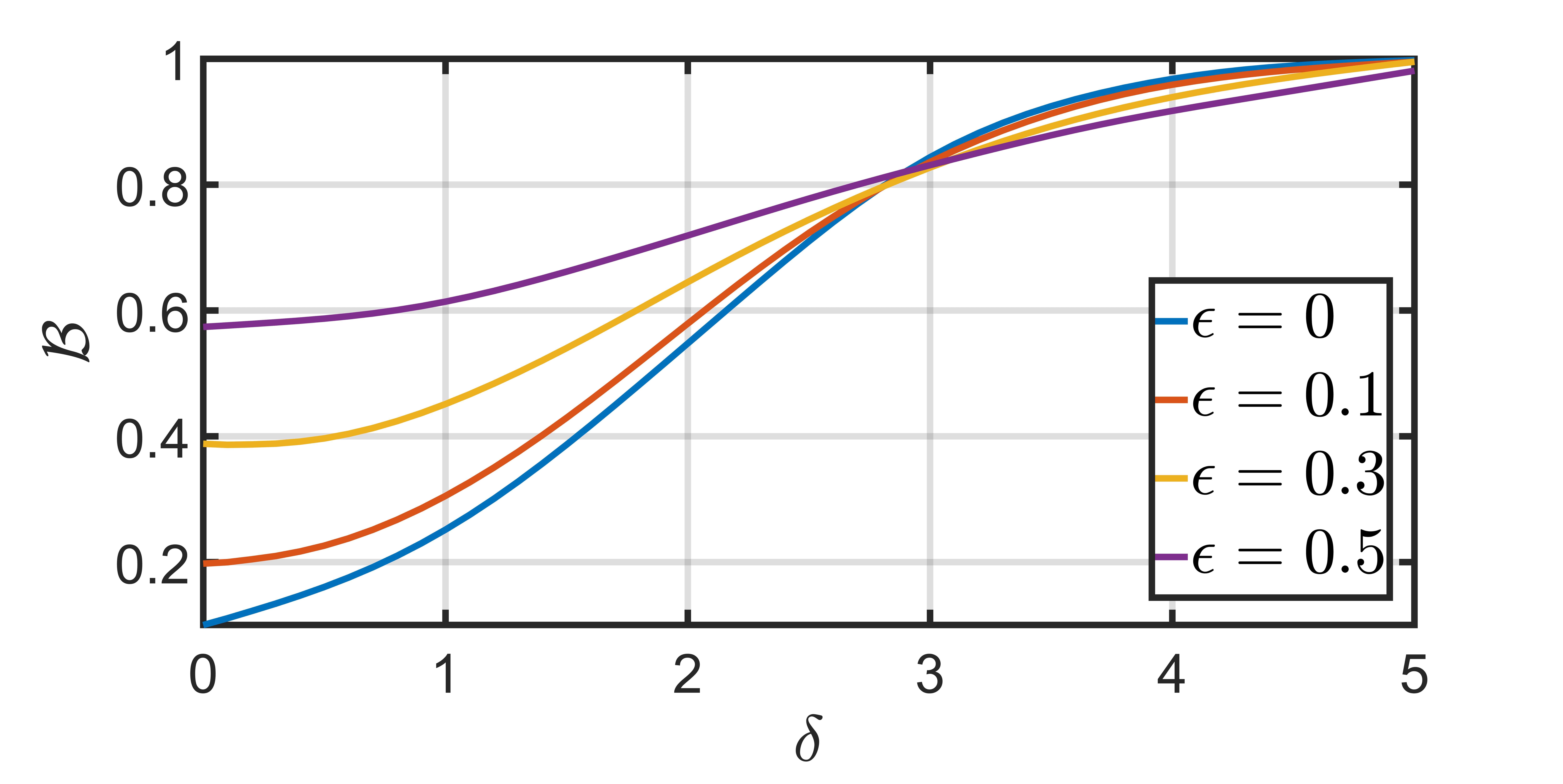}
  \caption{Comparison between the detection rate $\mathcal{B}$ of chi-squared tuning algorithm for an increasing amount of fault $\delta$ for the system without privacy-induced distortion ($\epsilon=0$) and system with privacy-induced distortions (with $\epsilon \in \{0.1,0.3,0.5\}$) and $\mathcal{A}^*=0.1$.}\label{detectionrate2}
\end{figure}

Next, as the second detection metric, the effect of privacy distortion on the detection rate of the chi-squared tuning procedure is depicted in Figure \ref{detectionrate2}. We consider deterministic additive constant fault $\delta_k=\delta$ with fault matrices $H=(0,1)^\top$ and $G=\mathbf{0}$. We apply privacy distortions with different distortion levels $\epsilon \in \{0,0.1,0.3,0.5\}$ and desired false alarm rate $\mathcal{A}^*=0.1$, where $\epsilon = 0$ indicates without privacy distortion. 
As can be seen in this figure, the detection rate of the anomaly detection algorithm $\mathcal{B}$, as defined in Section \Romannum{6}, is changing shape and flattening by increasing the distortion level $\epsilon$. For small $\delta$, the detection rate increases by increasing $\epsilon$, and for large $\delta$, the detection rate slightly decreases by increasing $\epsilon$. It might be mistakenly concluded from this figure that increasing the distortion level can improve the detection performance of the anomaly detector. On the other hand, Figure \ref{fig:approxfalsealarmrate} illustrates that by increasing the distortion level, the false alarm rate will increase. Therefore, to investigate the overall effect of privacy mechanisms on detection performances, the ROC curve is considered for analyzing the anomaly detector performance.

\begin{figure}[!htb]
  \centering
  \includegraphics[width=3.2in]{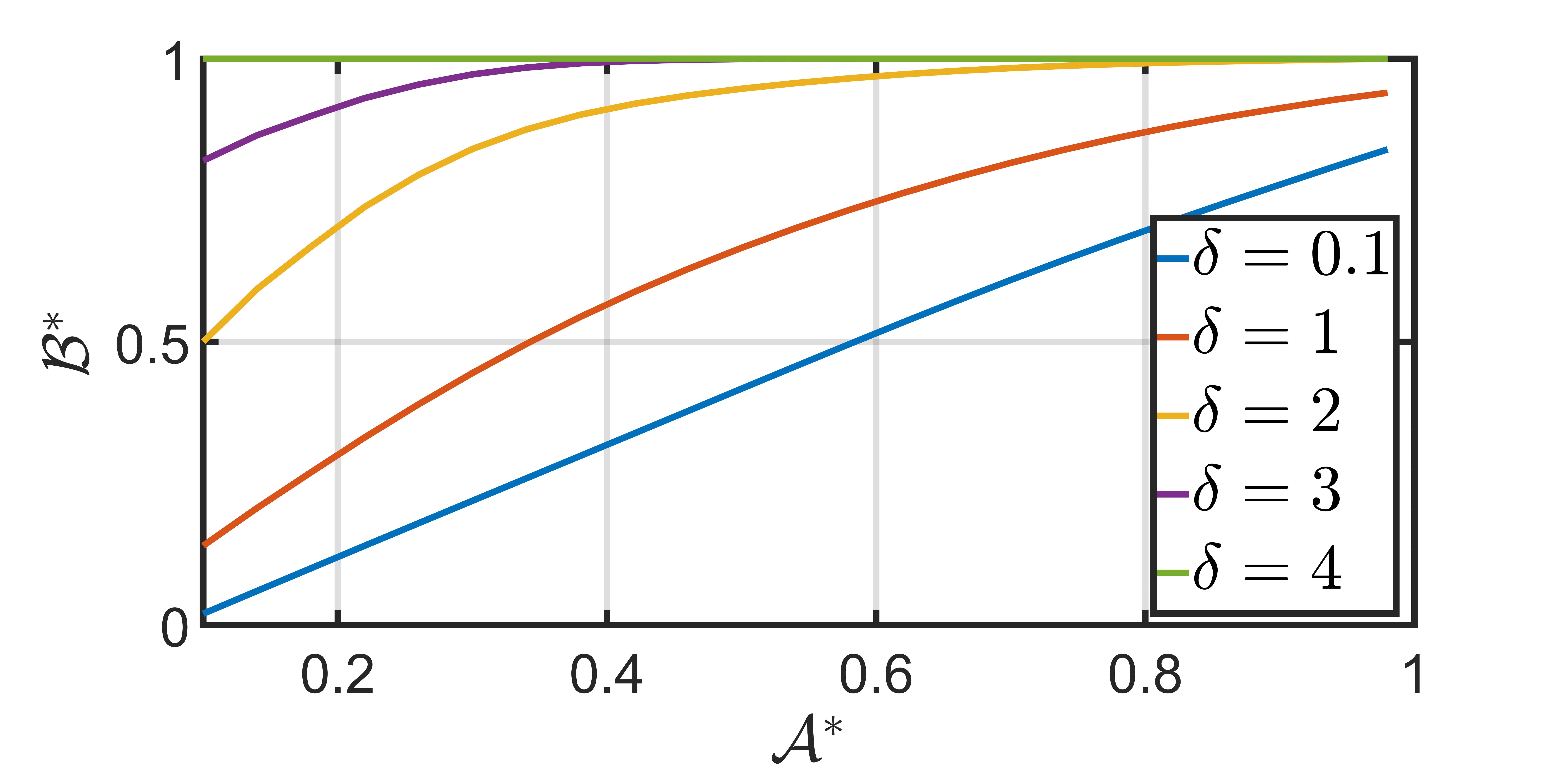}
  \caption{ROC curves of the detection rate in terms of desired false alarm rate without privacy distortion for different faults $\delta \in \{0.1,1,2,3,4\}$.}\label{ROCplot2}
\end{figure}

\begin{figure}[!htb]
  \centering
  \includegraphics[width=3.2in]{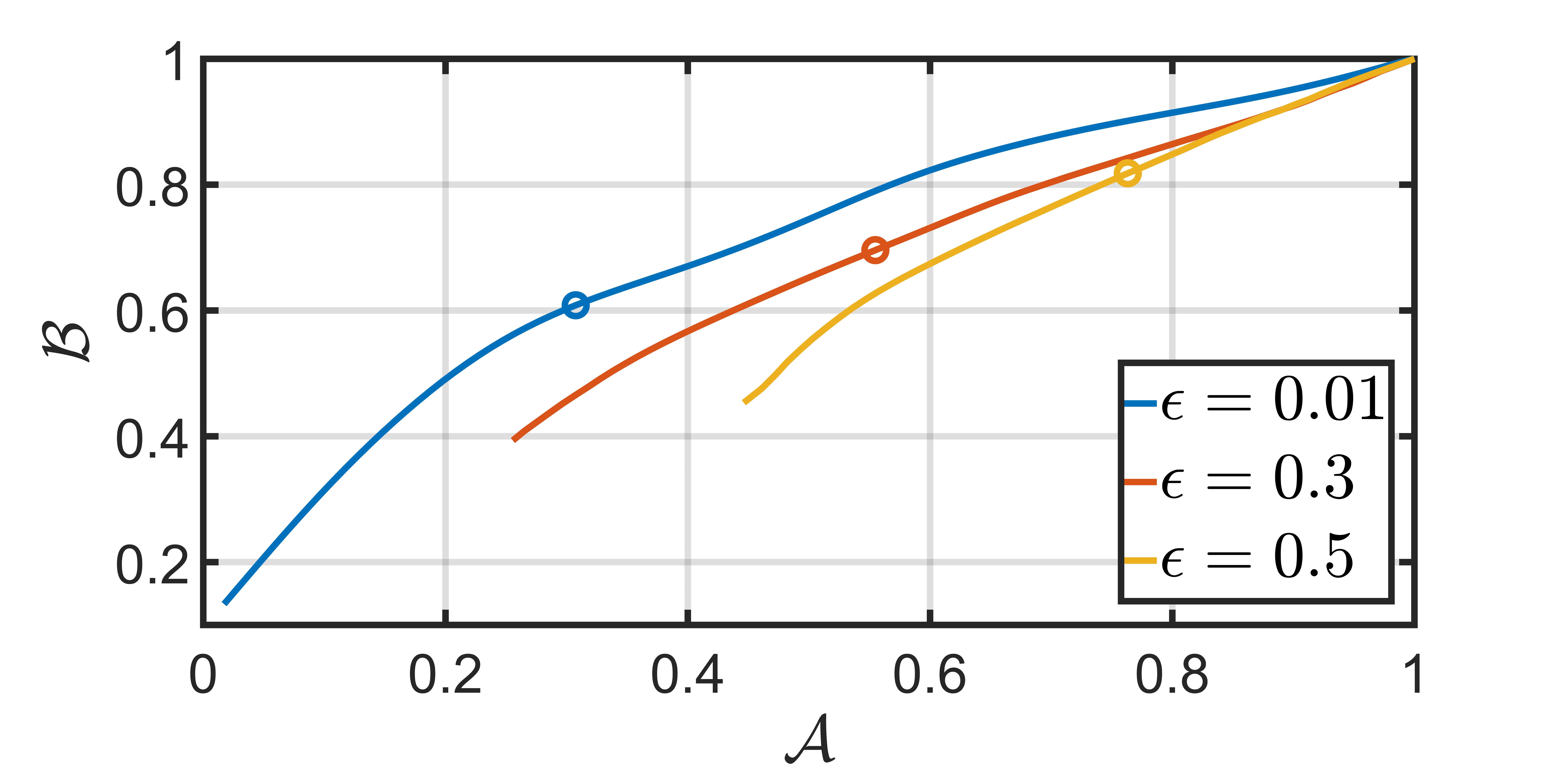}
  \caption{ROC curves of the detection rate in terms of the false alarm rate for different distortion levels $\epsilon \in \{0.01,0.3,0.5\}$, fixed fault $\delta=2$, and increasing $\mathcal{A}^*$.
  The circle markers on each line correspond to a certain value of the desired false alarm rate $\mathcal{A}^*=0.3$.}\label{ROCplot}
\end{figure}

The ROC shows how the detection rate changes as the false alarm rate increases. 
To characterize the ROC of the anomaly detection scheme, we find the detection rate (true positive) for a given fault $\delta$ as a function of the false alarm rate (false positive). Generally, in ROC analysis, tests are categorized based on the area under the ROC curve. The closer a ROC curve is to the upper left corner, the more efficient the test is. At first, in Figure \ref{ROCplot2}, we plot the ROC when there is no privacy distortion in the system for different amounts of fault $\delta \in \{0.1,1,2,3,4\}$. As we expected based on \eqref{detectionratewoprivacy}, by increasing the fault, the detection rate will be increased, and the ROC curve will be closer to the upper left corner, which illustrates the better performance of the anomaly detector. Then in Figure \ref{ROCplot}, the ROC is depicted for the distorted faulty system with a fixed fault $\delta=2$ and different distortion levels $\epsilon \in \{0.01,0.3,0.5\}$ by increasing the desired false alarm rate. As can be seen, for the fixed false alarm rate, the detection rate is decreased by increasing $\epsilon$. Also, for the fixed detection rate, the false alarm rate is increased by increasing $\epsilon$. In other words, by increasing $\epsilon$, the ROC curve will get farther from the upper left corner, which illustrates the worse performance of the anomaly detector. To further explore this figure, we marked with circle markers on all plots corresponding to a specific value of the desired false alarm rate $\mathcal{A}^*=0.3$. It can be seen that by increasing $\epsilon$, the $\mathcal{A}^*=0.3$ point moves toward the top right corner of the ROC, close to point $(1, 1)$ which shows a severely poor detection performance, equivalent to that of a detector that always identifies a residual as faulty, regardless of its real value. Therefore, this figure shows that increasing the privacy distortion will reduce the anomaly detector's performance.

\begin{figure}[!htb]
\centering
  \includegraphics[width=3.2in]{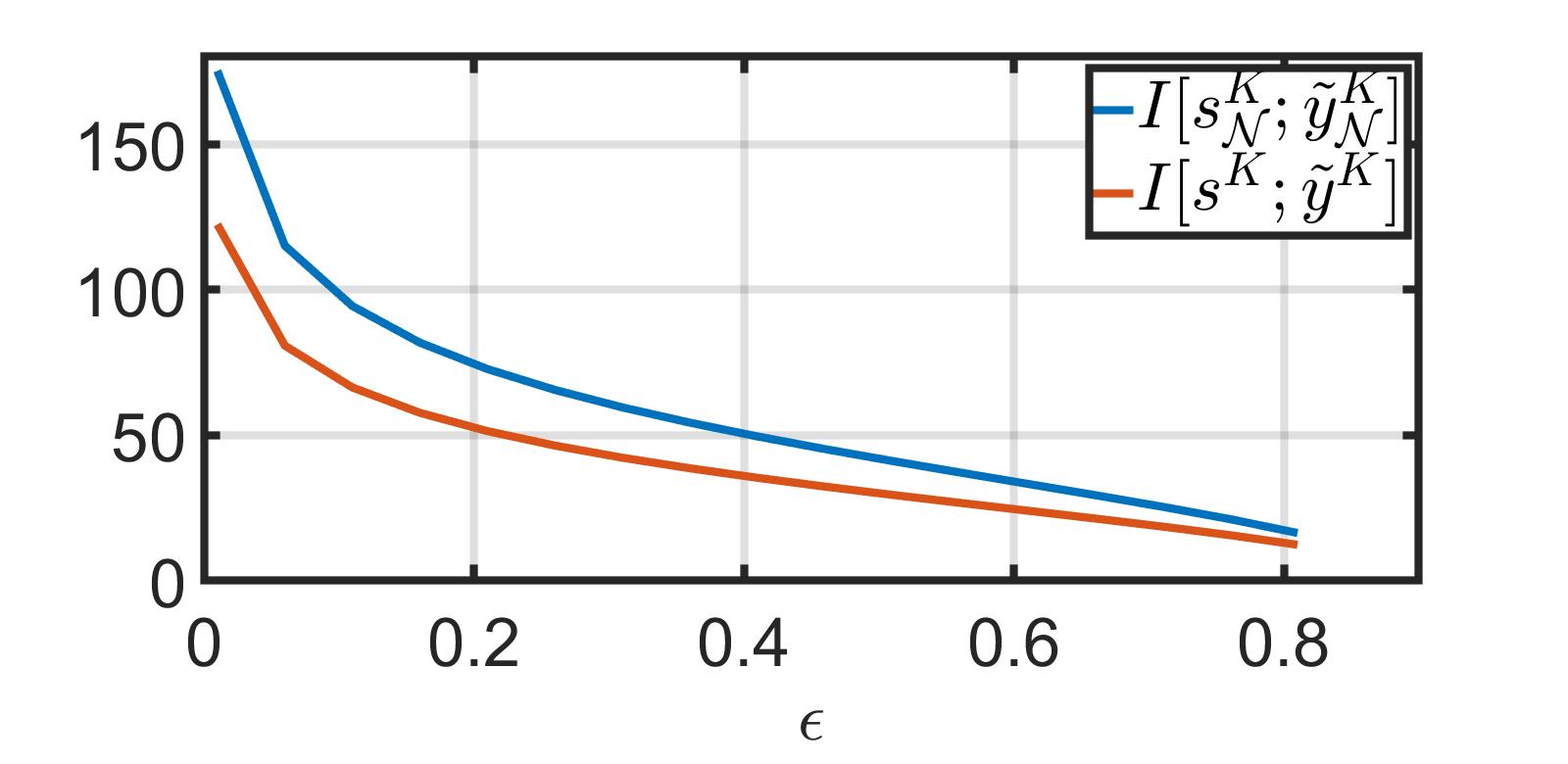}
  \caption{Comparison between the mutual information of Laplacian $s^K$ and $\tilde{y}^K$, $I[s^K;\tilde{y}^K]$, and the mutual information of Gaussian vectors $s^K_{\mathcal{N}}$ and $\tilde{y}^K_{\mathcal{N}}$, $I[s^K_{\mathcal{N}};\tilde{y}^K_{\mathcal{N}}]$, for increasing $\epsilon$ and desired false alarm rate $\mathcal{A}^*=0.1$.}
\label{laplacedistribution}
\end{figure}

In Figure \ref{laplacedistribution}, we show that, as discussed in Remark \ref{upperboundremark1} and \ref{upperboundremark}, for non-Gaussian random vectors $s^K$ and $\tilde{y}^K$ following joint log-concave distributions, their mutual information can be upper-bounded by affine functions of the mutual information of some Gaussian random vectors with the same mean and covariance matrices.
We consider $s^K$ and $\tilde{y}^K$ with joint Laplace distribution and solve the optimization problem in \eqref{eq:convex_optimization15} for $I[s^K_{\mathcal{N}};\tilde{y}^K_{\mathcal{N}}]-h[j^K]$ to find the optimal distorting parameters for an increasing amount of distortion level $\epsilon$. Since the maximum density \cite{bobkov2011entropy} of Gaussian random vectors and Laplace random vectors are the same for the same means and covariances, we can conclude that for Laplace variables $s^K_{{\mathcal{N}}^*}=s^K_{\mathcal{N}}$, $\tilde{y}^K_{{\mathcal{N}}^*}=\tilde{y}^K_{\mathcal{N}}$, and therefore $I[s^K_{{\mathcal{N}}^*};\tilde{y}^K_{{\mathcal{N}}^*}]-h[j^K]=I[s^K_{\mathcal{N}};\tilde{y}^K_{\mathcal{N}}]-h[j^K]$.
We calculate the optimal $I[s^K_{\mathcal{N}};\tilde{y}^K_{\mathcal{N}}]=I[s^K_{{\mathcal{N}}^*};\tilde{y}^K_{{\mathcal{N}}^*}]$ by solving the optimization problem \eqref{eq:convex_optimization15}. Then, we calculate $I[s^K;\tilde{y}^K]$ (see \cite{MImatlab} for the mutual information calculation), where $s^K$ and $y^K$ follow Laplace distributions with the same mean and covariance as $s^K_{\mathcal{N}}$ and $y^K_{\mathcal{N}}$, and $\tilde{y}^K=y^K +v^K$ (with the same distortion variables as the Gaussian case). As can be seen in Figure \ref{laplacedistribution}, $I[s^K;\tilde{y}^K]$ is even upper-bounded by $I[s^K_{\mathcal{N}};\tilde{y}^K_{\mathcal{N}}]$ without the constant terms $n$ and $C_n$. It implies that by solving the optimization problem for Gaussian variables, the information leakage for Laplace variables decreases for increasing $\epsilon$ and is upper-bounded by the information leakage for Gaussian variables. Furthermore, it is apparent that even if the Gaussian upper bounds are loose for these distributions of $s^K$ and $\tilde{y}^K$, the information leakage is still decreasing for increasing $\epsilon$. It shows that minimizing $I[s^K_{\mathcal{N}};\tilde{y}^K_{\mathcal{N}}]-h[j^K]$ or $I[s^K_{{\mathcal{N}}^*};\tilde{y}^K_{{\mathcal{N}}^*}]-h[j^K]$ effectively decreases the information leakage for any log-concave $s^K$ and $\tilde{y}^K$.


\section{Conclusions}
This paper presented an anomaly detection framework to maximize privacy for a class of stochastic dynamical systems. A detailed mathematical framework for synthesizing distorting mechanisms is given to minimize the information leakage induced by the use of public/unsecured communication networks. We proposed a class of linear Gaussian distorting mechanisms to randomize sensor measurements and input signals before transmission to prevent adversaries from accurately estimating the private part of the system state (a private performance output). Furthermore, for the class of systems under study, we thoroughly characterized information-theoretic metrics (mutual information and differential entropy) to quantify the information between private outputs and disclosed data for a class of worst-case eavesdropping adversaries.

Anomaly detection is an important problem that has been studied in a variety of applications. We proved that the privacy distorting mechanisms can increase the false alarm rate in the anomaly detector, resulting in reducing the performance of the anomaly detection scheme. Therefore, when designing the privacy distorting mechanism, we considered the trade-off between \emph{privacy} and \emph{anomaly detection performance}. We formulated the synthesis of the distorting mechanisms as a convex program, where we minimize the information leakage 
quantified using the mutual information between an arbitrarily large sequence of private outputs and the disclosed distorted data and the differential entropy of the sequence of input additive noise while considering a constraint on the amount of false alarm rate in the anomaly detector to limit the effect of privacy mechanisms on the anomaly detection algorithm. 
We presented simulation results to illustrate the performance of the proposed tools.
\bibliographystyle{IEEEtran}
\bibliography{TAC_Arxiv_210823}
\section{Appendix}
\subsection{Proof of Lemma 3}
$\tilde{z}_k$ and $q_k$ are defined as functions of variable $m_k$.
For the CDF of a function of a random variable, if $Y$ and $X$ are continuous random variables with $Y = g(X)$, the CDF of $Y$, $F_{Y}(y)$ can be characterized based on Probability Density Function (PDF) of $X$ as follows
\begin{equation}
    \begin{aligned}
F_{Y}(y) &=P_{Y}(Y \leq y) =P_{Y}(g(X) \leq y) \\
&=P_{X}(x \in \mathcal{X}: g(X) \leq y)
=\int_{\{x \in \mathcal{X}: g(X) \leq y\}} f_{X}(x) dx,
\end{aligned}
\end{equation}
where $f_{X}(x)$ and $\mathcal{X}$ are PDF and the sample space of $X$. Therefore, CDFs of $\tilde{z}_k$ and $q_k$ can be written as functions of the PDF of $m_k$ as below:
\begin{equation}\label{cdf1}
F_{\tilde{z}_{k}} (\alpha) =\int_{\{m_k: m_k^\top \Sigma'_k m_k \leq \alpha\}} f_{m_k}(\eta) d\eta,
\end{equation}
\begin{equation}\label{cdf2}
F_{q_{k}} (\alpha) =\int_{\{m_k: \beta_k m_k^\top {m_k} \leq \alpha\}} f_{m_k}(\eta) d\eta.
\end{equation}
From $ m_k^\top \Sigma'_k m_k \le \beta m_k^\top m_k$, we can conclude that
\begin{equation}
    {\{m_k: \beta_k m_k^\top m_k \leq \alpha\}}  \subseteq {\{m_k: m_k^\top \Sigma'_k m_k \leq \alpha\}},\label{subset}
\end{equation}
which means that the ellipsoid $\{\beta_k m_k^\top m_k = \alpha\}$ is a subset of ellipsoid $\{m_k^\top \Sigma'_k m_k = \alpha\}$ (see \cite{Boyd2004} for proof and details). \\
From \eqref{subset} and the fact that $f_{m_k}(\eta)$ is always non-negative, we deduce that
\begin{equation}\label{finalcdf}
    \int_{\{m_k: m_k^\top \Sigma'_k m_k \leq \alpha\}} f_{m_k}(\eta) d\eta \ge \int_{\{m_k: \beta_k m_k^\top m_k \leq \alpha\}} f_{m_k}(\eta) d\eta.
\end{equation}
According to \eqref{cdf1} and \eqref{cdf2}, \eqref{finalcdf} is equivalent to $ F_{\tilde{z}_{k}} (\alpha) \ge F_{q_k} (\alpha)$.
\hfill $\blacksquare$
\subsection{Proof of Lemma 4}
By multiplying $\tilde{\Sigma}^{-1/2}_k$ to the right and $(\tilde{\Sigma}^{-1/2}_k)^\top$ to the left side of $(\tilde{\Sigma}^{1/2}_k)^\top \Sigma^{-1} \tilde{\Sigma}^{1/2}_k \le \beta^* {I}_{n_y}$, it will be converted to
\begin{equation}
    \Sigma^{-1} \le \beta^* (\tilde{\Sigma}_k^\top \tilde{\Sigma}_k)^{-1/2} =\beta^* \tilde{\Sigma}_k^{-1}.\label{eq:lemma42}
\end{equation}
Since $\Sigma$ and $\tilde{\Sigma}_k$ are positive definite matrices, \eqref{eq:lemma42} can be written as \eqref{eq:lemma41} which is a convex function of design variables $\Sigma^v_k$ and $\Sigma^j_k$.
\hfill $\blacksquare$
\subsection{Proof of Proposition 1}
First, we need to prove that for initial state $x_1$ and disturbances $t_k$ and $w_k$ following non-Gaussian log-concave distributions, and $j^K$ and $v^K$ following Gaussian distributions, $s^K$ and $\tilde{y}^K$ follow log-concave distributions. The log-concavity of distribution is preserved under affine transformations (see \cite{dharmadhikari1988unimodality}, Lemma 2.1). Besides, based on Hoggar’s theorem, the sum of two independent log-concave random variables is itself log-concave \cite{hoggar1974chromatic,johnson2006preservation}. Therefore, based on the system dynamics \eqref{eq1}, distorting mechanism \eqref{eq2}, the independency between initial state and disturbances, and also independency between $v^K$, $j^K$, and $y^K$, we can conclude that $y^K$, $s^K$, and $\tilde{y}^K$, follow log-concave distributions.

Mutual information $I\left[s^K;\tilde{y}^K\right]$ can be written in terms of differential entropies as $I[s^K;\tilde{y}^K] = h[s^K] + h[\tilde{y}^K] - h[s^K,\tilde{y}^K]$. Moreover, because Gaussian distributions maximize entropy for a given covariance matrix \cite{Cover}, we can write:
\begin{equation}
    \left\{\begin{aligned}
h[s^K] &\le h[s^K_{\mathcal{N}}],\\
h[\tilde{y}^K] &\le h[\tilde{y}^K_{\mathcal{N}}].
\end{aligned}\right. \label{inequalityentropy}
\end{equation}
The authors in \cite{marsiglietti2018lower} prove that the entropy of random vectors following log-concave distributions is lower bounded as:
\begin{equation}
    h[s^K,\tilde{y}^K] \ge \frac{n}{2} \log \frac{\left|\Sigma_K^{\tilde{y},s}\right|^{1 / n}}{c(n)},\label{entropylower}
\end{equation}
where $n=K(n_s + n_y)$ and $c(n)=\frac{e^{2} n^{2}}{4 \sqrt{2}(n+2)}$. Expanding the lower bound in \eqref{entropylower}, we obtain
\begin{align}\label{lowerboundformula}
   \frac{n}{2} \log \frac{\left|\Sigma_K^{\tilde{y},s}\right|^{1 / n}}{c(n)} &=\frac{n}{2}\left(\frac{1}{n} \log \operatorname{det} \Sigma_K^{\tilde{y},s}-\log c(n)\right)\nonumber \\
   &=\frac{1}{2} \log \operatorname{det} \Sigma_K^{\tilde{y},s} - \frac{n}{2} \log c(n).
\end{align}
Let $(s^K_{\mathcal{N}},\tilde{y}^K_{\mathcal{N}})$ be jointly Gaussian with covariance $\Sigma_K^{\tilde{y},s}$. Then, we can write its differential entropy as $h[s^K_{\mathcal{N}},\tilde{y}^K_{\mathcal{N}}] = \frac{1}{2}\log \det \left( \Sigma_K^{\tilde{y},s} \right) + \frac{n}{2} + \frac{n}{2}\log(2\pi)$ \cite{Cover}. Therefore, using the latter formula for $h[s^K_{\mathcal{N}},\tilde{y}^K_{\mathcal{N}}]$ and \eqref{lowerboundformula}, we can write the following
\begin{equation}
     \frac{n}{2} \log \frac{\left|\Sigma_K^{s,\tilde{y}}\right|^{1 / n}}{c(n)}= h[s^K_{\mathcal{N}},\tilde{y}^K_{\mathcal{N}}] - \frac{n}{2} - \frac{n}{2}\log(2\pi) - \frac{n}{2} \log c(n).\label{entropylower2}
\end{equation}
Combining \eqref{inequalityentropy}, \eqref{entropylower}, and \eqref{entropylower2}, we can finally conclude that
\begin{align}
       I[s^K;\tilde{y}^K] &\le h[s^K_{\mathcal{N}}] + h[\tilde{y}^K_{\mathcal{N}}] - h[s^K_{\mathcal{N}},\tilde{y}^K_{\mathcal{N}}]\nonumber\\ &+\frac{n}{2} + \frac{n}{2}\log(2\pi)+ \frac{n}{2} \log c(n),
\end{align}
which is equivalent to \eqref{eqmutualinfupper1}.\hfill $\blacksquare$
\subsection{Proof of Proposition 2}
Mutual information $I\left[s^K;\tilde{y}^K \right]$ can be written in terms of differential entropies as $I[s^K;\tilde{y}^K] = h[s^K] + h[\tilde{y}^K] - h[s^K,\tilde{y}^K]$. Besides, it has been proved in Appendix C that $y^K$, $s^K$, and $\tilde{y}^K$, follow log-concave distributions.  Then, from the inequality on the entropy of log-concave distributed random vectors, we can conclude that (see Theorem $I.1$ in \cite{bobkov2011entropy}):
\begin{equation}
    \left\{\begin{aligned}
\frac{1}{Kn_s} h[s^K] &\le \frac{1}{Kn_s}  h[s^K_{{\mathcal{N}}^*}] + \frac{1}{2},\\
\frac{1}{Kn_y} h[\tilde{y}^K] &\le \frac{1}{Kn_y} h[\tilde{y}^K_{{\mathcal{N}}^*}] + \frac{1}{2},\\
\frac{1}{n} h[s^K,\tilde{y}^K] &\ge \frac{1}{n} h[s^K_{{\mathcal{N}}^*},\tilde{y}^K_{{\mathcal{N}}^*}] - \frac{1}{2}.
\end{aligned}\right. \label{inequalityentropy2}
\end{equation}
Combining all the inequalities in \eqref{inequalityentropy2}, we have
\begin{align}
       \frac{1}{n} I[s^K;\tilde{y}^K] &\le \frac{1}{n}(h[s^K_{{\mathcal{N}}^*}] + h[\tilde{y}^K_{{\mathcal{N}}^*}] - h[s^K_{{\mathcal{N}}^*},\tilde{y}^K_{{\mathcal{N}}^*}]) + 1,
\end{align}
which is equivalent to \eqref{eqmutualinfupper}.\hfill $\blacksquare$
\subsection{Proof of $\tilde{z}_{k}$ approximately following a gamma distribution}
At first, we prove that defining $m_k$ as a standard normal variable $m_k \sim \mathcal{N}(\mathbf{0},I_{n})$ and matrix $Q \in \mathbb{R}^{n \times n}$ with $n \in \mathbb{R}^+$, $Q' =  m_k^\top Q m_k$ approximately follows a gamma distribution.
$Q' =  m_k^\top Q m_k$ is a weighted sum of chi-squared distributions with one degree of freedom ($\chi^2_1$) as $Q' \sim \sum\limits_{i = 1}^{{n}} \lambda_i \chi^2_1$ \cite{feiveson1968distribution}, where $\lambda_i$, $i \in \{1,\ldots,n\}$, are the eigenvalues of the matrix $Q$.
For each term of the sum $\lambda_i \chi^2_1$ we have \cite{stacy1962generalization}:
\begin{equation}
    \lambda_i \chi^2_1 \sim \Gamma(K_i = \frac{1}{2}, \theta_i = 2 \lambda_i).
\end{equation}
where $\Gamma(K_i = \frac{1}{2}, \theta_i = 2 \lambda_i)$ is a gamma variable with shape parameter $K_i = \frac{1}{2}$ and scale parameter $\theta_i = 2 \lambda_i$. Therefore, $Q'  \sim \sum\limits_{i = 1}^{{n}} \Gamma(K_i = \frac{1}{2}, \theta_i = 2 \lambda_i)$ is the sum of gamma variables. Based on the Welch-Satterthwaite approximation \cite{satterthwaite1946approximate,welch1947generalization}, the summation of gamma distributions $\sum\limits_{i = 1}^{{n}} \Gamma(K_i,\theta_i)$ approximately follows a gamma distribution with shape $K_{sum}=\frac{\left(\sum_{i} \theta_{i} K_{i}\right)^{2}}{\sum_{i} \theta_{i}^{2} K_{i}}$ and scale $\theta_{sum}=\frac{\sum \theta_{i} K_{i}}{K_{sum}}$.
Consequently, $Q'$ follows a gamma distribution $Q' \sim \Gamma(K,\theta)$ with shape $K=\frac{\left(\sum_{i=1}^{n} \lambda_i\right)^{2}}{2\sum_{i=1}^{n}\left(\lambda_{i}\right)^{2}}$ and scale $\theta=2\frac{\sum_{i=1}^{n}\left(\lambda_{i}\right)^{2}}{\left(\sum_{i=1}^{n} \lambda_{i}\right)}$.
The accuracy of this approximation is proved in \cite{feiveson1968distribution}.

Hence, we can conclude that $\tilde{z}_{k} = m_k^\top \Sigma'_k m_k$ approximately follows a gamma distribution $\tilde{z}_{k} \sim \Gamma(K_{z}^{k},\theta_{z}^{k})$ with:
\begin{eqnarray}
&K_{z}^{k}=\frac{\left(\sum_{i=1}^{n_{y}} \lambda_{i,k}\right)^{2}}{2 \sum_{i=1}^{n_{y}}\left(\lambda_{i,k}\right)^{2}},\\ &\theta_{z}^{k}=2\frac{\sum_{i=1}^{n_{y}}\left(\lambda_{i,k}\right)^{2}}{\left(\sum_{i=1}^{n_{y}} \lambda_{i,k}\right)},
\end{eqnarray}
where $\lambda_{i,k}$, $i \in \{1,\hdots,n_y \}$ are the eigenvalues of the matrix $\Sigma'_k = (\tilde{\Sigma}^{1/2}_k)^\top \Sigma^{-1} \tilde{\Sigma}^{1/2}_k$. 

It is shown in \cite{feiveson1968distribution} that the mean of $\lambda_{i,k}$, which in our case are the mean of eigenvalues of ${\Sigma}'_k$, does not affect the accuracy of this approximation. Therefore, for added noises with relatively large covariances ($\Sigma^v_k$ and $\Sigma^j_k$), matrix $\Sigma'_k$ would be larger, which lead to larger eigenvalues of ${\Sigma}'_k$; but it does not affect the accuracy of this approximation. \hfill $\blacksquare$

\end{document}